\documentclass[reprint,
superscriptaddress,
nofootinbib,
nobibnotes,
 amsmath,amssymb,
 aps,
longbibliography
]{revtex4-2}

\usepackage{verbatim}
\usepackage{graphicx}
\usepackage{dcolumn}
\usepackage{bm}
\usepackage{color}

\usepackage[hidelinks]{hyperref}
\begin{document}

\title{Stochastic Inflation in General Relativity}

\author{Yoann L. Launay}
 
  \email{yoann.launay@outlook.com}
 \affiliation{Centre for Theoretical Cosmology, Department of Applied Mathematics and Theoretical Physics,
University of Cambridge, Wilberforce Road, Cambridge CB3 0WA, United Kingdom}

\author{Gerasimos I. Rigopoulos}
\email{gerasimos.rigopoulos@ncl.ac.uk}
\affiliation{School of Mathematics, Statistics and Physics,  Newcastle University, Newcastle upon Tyne, NE1 7RU, United Kingdom}

\author{E. Paul S. Shellard}
\email{epss@damtp.cam.ac.uk}
\affiliation{Centre for Theoretical Cosmology, Department of Applied Mathematics and Theoretical Physics,
University of Cambridge, Wilberforce Road, Cambridge CB3 0WA, United Kingdom}

\begin{abstract}
We provide a formulation of Stochastic Inflation in full general relativity that goes beyond the slow-roll and separate universe approximations. We show how gauge invariant Langevin source terms can be obtained for the complete set of Einstein equations in their ADM formulation by providing a recipe for coarse-graining the spacetime in any small gauge. These stochastic source terms are defined in terms of the only dynamical scalar degree of freedom in single-field inflation and all depend simply on the first two time derivatives of the coarse-graining window function, on the gauge-invariant mode functions that satisfy the Mukhanov-Sasaki evolution equation, and on the slow-roll parameters. It is shown that this reasoning can also be applied to include gravitons as stochastic sources, thus enabling the study of all relevant degrees of freedom of general relativity for inflation. We validate the efficacy of these Langevin dynamics directly using an example in uniform field gauge, obtaining the stochastic e-fold number in the long wavelength limit without the need for a first-passage-time analysis. As well as investigating the most commonly used gauges in cosmological perturbation theory, we also derive stochastic source terms for the coarse-grained BSSN formulation of Einstein's equations, which enables a well-posed implementation for 3+1 numerical relativity simulations.  
\end{abstract}
\maketitle

\section{Introduction \label{sec:intro}}

Inflation theory was postulated more than 40 years ago as an explanation for the apparently fine-tuned initial conditions of the Hot Big Bang \cite{starobinsky80,guth_1981,Linde82}. The proposal gained traction as it also offers a natural mechanism for generating the initial density inhomogeneities \cite{Mukhanov1981,Hawking1982,Starobinsky1982pert,Guth82pert,HawkingMoss1982,Bardeen83} which in later stages of cosmic history led to the formation of cosmic structure via gravitational instability. These density fluctuations are directly observable in the Cosmic microwave background and their two-point statistics have been measured to very high precision \cite{PlanckInflation}. 

Inflation also predicts the existence of a cosmological gravitational wave background \cite{Rubakov82}, yet to be detected,  as well as the possible existence of non-zero higher-order spatial correlation functions in the cosmological fluid \cite{Maldacena_2003}. The latter's amplitude, and the amount of non-gaussianity more generally, depend more heavily on the specifics of the inflationary model. They could be detectable in the cosmic microwave background (CMB) and cosmological structures and even lead to the formation of primordial black holes. If detected, any such features would open up remarkable windows into the early universe.

Central to the inflationary origin scenario is the assumption that our universe originated from quantum processes. This underscores the necessity to combine quantum mechanics and gravity to make precise predictions for inflationary models of the early universe. Although a theory of quantum gravity is lacking, despite decades of efforts \cite{WdW,Hawking78, WaldQFTCS, polchinski1998, Rovelli90, TriangularQG}, there are regimes where predictions can still be made by using techniques from Quantum Field Theory on Curved Spacetime (QFTCS) \cite{WaldQFTCS} or by constructing Effective Field Theories (EFTs), which have made continuous advancements in cosmology, inspired by the latter's success in flat space \cite{polchinski1999effective}. Abandoning pretenses of completeness, an EFT establishes a region of validity, normally bounded by ultraviolet (UV) and/or infrared (IR) cutoffs, and the narrative of theoretical physics is implicitly about pushing these cutoffs to their limits.

Not long after the original concept of inflation was introduced, Stochastic Inflation (SI) was formulated in  \cite{Starobinsky_stochastic_1988}, was used as a non-perturbative methodology for light scalars in de Sitter in \cite{starobinsky_equilibrium_1994}, and was later adopted as a special kind of EFT for inflationary physics, as part of a semiclassical stochastic gravity approach \cite{verdaguer_stochastic_2007}. Stochastic Inflation therefore, can be thought of as an EFT for light scalar fields in quasi-De Sitter spacetimes that sets its UV cutoff at the point where quantum operators governing the field perturbations can be described as classical stochastic variables and super-Hubble correlators can automatically exhibit decoherence, a distinctive feature of such systems. In most inflation models, this transition occurs when Fourier modes cross the Hubble radius and their transition into the EFT's range of validity is modelled by the action of a small stochastic noise perturbing the long wavelength fields described by General Relativity - in a sense, Stochastic Inflation is a combination of deterministic evolution with a continuous reset of initial conditions by small stochastic amounts coming from the influx of new modes which cross the UV threshold. The stochastic equations can then be used to derive non-perturbative results such as resummations of classical loops which are IR divergent in QFTCS \cite{moss_effective_2017,Gorbenko19,cespedes23}.

In this work, as in most others on the topic, the sub-Hubble scales will be described by linear cosmological perturbation theory (CPT); although higher order perturbation theory can in principle be used for these sub-Hubble scales \cite{cohen_stochastic_2021}, we will not undertake such an endeavour here. Stochastic Inflation therefore, in its current formulation, sacrifices non-linearities arising in the sub-Hubble regime in favor of \emph{a fully non-perturbative theory in the super-Hubble regime}, where QFTCS might fail as it has already been reported \cite{Gorbenko19}. It is therefore particularly suited for circumstances where non-linear evolution on super-Hubble scales, and possibly non-perturbative correlators or even full probability density functions (PDFs), are the relevant quantities to compute.

Although able to provide non-perturbative quantities, Stochastic Inflation as usually formulated comes with several approximations that simplify the complex dynamics of General Relativity, most notably relying on the Separate Universe Approximation (SUA) \cite{salopek_nonlinear_1990,salopek_stochastic_1991,wands_new_2000} and therefore altogether dropping the dynamics of some degrees of freedom of the gravitational field. To find the spectrum of the stochastic perturbations, it has also been claimed to require a specific choice for the time-slicing to ensure consistency with QFTCS, namely using a uniform e-fold time slicing \cite{Finelli09,Finelli10,pattison_stochastic_2019,artigas_hamiltonian_2022}. However, there are formulations where such restrictions are not required \cite{rigopoulos_non-linear_2005,cruces_stochastic_2022}, which appears to be linked to an ongoing discussion about the appropriate inclusion of the momentum constraint \cite{prokopec_ensuremathdeltan_2021,cruces_stochastic_2022}. Furthermore, obtaining the observationally relevant curvature perturbation then requires a ``first passage time'' analysis (FPTA) to determine a stochastic number of e-folds for a specified point of the scalar potential to be reached \cite{vennin_correlation_2015}.            

In this article, we endeavour to demonstrate that it is possible to generalize the stochastic inflation equations by making the linear treatment on sub-Hubble scales the sole assumption, thereby removing the need for all the aforementioned approximations. More specifically, we formulate equations for Stochastic Inflation within full General Relativity in its ADM formulation, retaining all the variables describing the gravitational field and the freedom to choose the time-slicing and the spatial coordinates within the 3D time-slices. 

To achieve this, we will proceed as follows: After recalling the ADM formulation and its linearized version in Sec.\ \ref{sec:secII}, we survey a range of commonly used gauges and gauge invariant variables in Sec.\ \ref{sec:secIII}, where we also provide explicit expressions for determining the spacetime foliation and the dynamical variables from knowledge of the gauge invariant comoving curvature perturbation $\mathcal{R}$. Our procedure for coarse-graining and obtaining stochastic source terms for the full set of ADM equations is explained in Sec.\ \ref{sec:secIV}. These stochastic sources are expressed in terms of $\mathcal{R}$ and are found to be identical in all the small gauges we examine. We therefore postulate that they represent the complete small gauge invariant stochastic continuation of the Einstein equations with linearized source terms, noting that they satisfy linear perturbation theory by construction in any choice of gauge/foliation. Sec. \ref{sec:secIVF} extends the result to include tensorial modes sources. As a short application, we present and solve stochastic equations for $\Delta{\cal N}$ in Sec.\ \ref{sec:secV} in a toy model exhibiting a short phase of ultra slow-roll, directly obtaining the PDF of the curvature perturbation without the need for a first-passage-time analysis.

\begin{center}
    *****
\end{center}

\textbf{Notations:} Units are such that $\hbar = c = 1$. The Planck mass is written $M_{Pl}= (8\pi G)^{-\frac{1}{2}}$. Quantum operators will carry  $\hat{()}$ while random variables will be bold symbols. We adopt the following Fourier transform convention: ${\cal F}^{-1}\{\cdot\} = \int \frac{d^3\vec{k}}{\sqrt{2\pi}^3} (\cdot)e^{-i\vec{k}\cdot\vec{x}}$. When it is needed, a background field with respect to cosmological perturbation theory will carry a '$b$' subscript. $\dot{()}$ will be time coordinate partial derivatives if not explicitly total.

\section{ADM equations and scalar linear perturbations \label{sec:secII}}

In this section, we review the ADM formulation of the equations of General Relativity and their linearized perturbations around a homogeneous and isotropic background. For further purposes that will become clear further ahead, all equations are written as left-hand-sided.

\subsection{ADM formulation of Einstein's equations \label{sec:secIIA1}}
For an unknown 4-dimensional metric tensor with components $g_{\mu\nu}$ describing a spacetime with Ricci scalar $R$ and 'matter' Lagrangian density ${\cal L}_m$, the Einstein-Hilbert action with vanishing Gibbons-Hawking-York boundary terms \cite{GHY_Y,GHY_GH} is
\begin{equation}
S =  \frac{M_{Pl}^2}{2} \int d^4 x \sqrt{- g} \, R +  \int d^4 x \, {\cal L}_m.
\end{equation}
When varied with respect to the metric $g_{\mu\nu}$ it gives Einstein's equations
\begin{equation}
    G_{\mu\nu}-M_{Pl}^{-2}\, T_{\mu\nu} = 0,
\end{equation}
where $T^{\mu\nu} = -\frac{2}{\sqrt{-g}}\frac{\delta {\cal L}_m}{\delta g_{\mu\nu}}$ such that $\nabla_{\mu}T^{\mu\nu} = 0$ by construction. In this work, we will study a single $\phi$ scalar field fluid described by
\begin{equation}
    {\cal L}_m = -\sqrt{-g}\left[\frac{1}{2}\, g^{\mu\nu} \nabla_{\mu}\phi\nabla_{\nu}\phi+ V(\phi)\right].
\end{equation}

In what follows we will use the common notations for the 3+1 splitting of ADM formalism \cite{ADM}, letting $n$ and $P$ be the projectors along the normal of the 3-space submanifold and the submanifold itself respectively. The associated covariant derivative component on the 3-space submanifold will be denoted by a vertical bar $()_{|i}$.

We start from the ADM parameterization of the metric, assuming the existence of a foliation of spacetime
\begin{equation}
    d s^2=-\alpha^2 d t^2+\gamma_{i j}\left(d x^i-\beta^i d t\right)\left(d x^j -\beta^j d t\right).
    \label{eq:ADMmetric}
\end{equation}
The Einstein-Hilbert action can be re-expressed to reflect this $3+1$ splitting 
\begin{equation}
\begin{aligned}
S =  \frac{M_{Pl}^2}{2} \int d t d^3 x \, \alpha \sqrt{\gamma}\left[{ }^{3} R-K^2+K_{i j} K^{i j}\right]& \\
 +\int d t d^3 x \, \alpha \sqrt{\gamma}\left[\frac{1}{2} \Pi^2-\frac{1}{2} \partial_i \phi \partial^i \phi-V(\phi)\right], &
\end{aligned}
    \label{eq:EinstHilbert}
\end{equation}
where ${}^3R$ the Ricci scalar of the spatial metric $\gamma_{ij}$, and $\Pi$ is the conjugate momentum
\begin{equation}
\Pi= n^\mu \partial_\mu \phi=\frac{1}{\alpha}\left(\dot{\phi}+\beta^i \partial_i \phi\right),
    \label{eq:Mom}
\end{equation}
and $K$ is the extrinsic curvature
\begin{equation}
K_{ij} \equiv n_{i|j} = -\frac{1}{2\alpha}(\gamma_{i j,0}+\beta_{i|j}+\beta_{j|i}),
    \label{eq:eqK}
\end{equation}
which is usually split into its trace and traceless components $K$ and $\tilde{K}_{ij}$ respectively.

Variation of  (\ref{eq:EinstHilbert}) with respect to $\phi$  yields the  field evolution equation
\begin{equation}
    \frac{1}{\alpha}\left(\dot{\Pi}+\beta^i \Pi_{\mid i}\right)-K \Pi-\frac{\alpha^{\mid i}}{\alpha} \phi_{\mid i}-\phi_{\mid i}^{\mid i}+\frac{d V}{d \phi}=0,
    \label{eq:Pieq}
\end{equation}
while variation with respect to $\gamma_{i j}$  yields the gravitational field's evolution equations 
\begin{equation}
\left \{
\begin{aligned}
 \dot{K}+\beta^i K_{, i}+\alpha^{\mid i}{ }_{\mid i}-\alpha\left({ }^{3} R+K^2\right)&\\
 - M_{Pl}^{-2} \alpha\left(\frac{1}{2} S-\frac{3}{2} \rho\right) = 0, & \\
  \dot{\tilde{K}}_{ij}+2\alpha\tilde{K}_{il}{\tilde{K}}^l{ }_j+\beta^k \tilde{K}_{ij \mid k}-2\beta_i{ }^{\mid k} \tilde{K}_{jk}+ \alpha_{\mid i\mid j}&\\
  -\frac{1}{3} \alpha^{\mid k}{ }_{\mid k} \delta_{ij} 
-\alpha\left({ }^3\tilde{R}_{ij}+\frac{1}{3}K \tilde{K}_{ij}\right) + M_{Pl}^{-2} \alpha \tilde{S}_{ij}=0, &
\end{aligned}\right .
    \label{eq:Keq}
\end{equation}
where the $3+1$ components of the energy-momentum tensor are the energy and momentum densities
\begin{equation}
\left\{
\begin{aligned}
\rho & =n_\mu n_\mu T^{\mu v}=\alpha^2 T^{00}=\frac{1}{2} \Pi^2+\frac{1}{2} \partial_i \phi \partial^i \phi+V(\phi), \\
{\cal { J}}_i & = -n_{\mu}P_{ \nu i} T_{\mu \nu} = \alpha T^{0}{}_i = -\Pi \partial_i\phi,
\end{aligned}\right .
\label{eq:densities}
\end{equation}
and the stress tensor $S_{i j}=P_{ \mu i}P_{ \nu j} T^{\mu \nu} = T_{ij}$, decomposed as
\begin{equation}
    \left \{
    \begin{aligned}
        S & \equiv{ }^{3} g^{i j} S_{i j}=\frac{3}{2} \Pi^2-\frac{1}{2} \partial_k \phi \partial^k \phi-3 V(\phi), \\
\tilde{S}_{i j}& \equiv S_{ij}-\frac{1}{3}{ }^{3} g^{i j}S =  \frac{1}{2}\left(\partial_i \phi \partial_j \phi-\frac{1}{3} \partial_k \phi \partial^k \phi \,\delta_{i j}\right) .
    \end{aligned} \right .
        \label{eq:SETdef}
\end{equation}
Finally, variation with respect to $\alpha$ and $\beta^i$ leads to two constraints 
\begin{equation}
\left \{
\begin{aligned}
 { }^{3}R+\frac{2}{3} K^2-\tilde{K}_{i j} \tilde{K}^{i j}-2 M_{Pl}^{-2} \rho = 0, &\\
 \tilde{K}^j{}_{i \mid j}-\frac{2}{3} K_{\mid i}- M_{Pl}^{-2} {\cal { J}}_i=0, & 
\end{aligned}\right .
    \label{eq:cons_eq}
\end{equation}
usually referred to as the Hamiltonian and Momentum constraints respectively. 

\subsection{Scalar Linear equations \label{sec:secIIB}}

This section reviews the linearised version of the ADM formalism of scalar perturbation theory without making any gauge choice \cite{PaulADM}. The linear equations presented below match exactly the non-linear left-hand sides (LHS) of the previous subsection. Since the standard Einstein and ADM systems of equations are equivalent up to substitutions and factors, the equations below are also equivalent to the more commonly adopted formulations of cosmological perturbation theory.

We can incorporate general scalar perturbations through the line element
\begin{equation}
\begin{aligned}
    d s^2=-\alpha_b^2(1+2 \Psi) d t^2+2 a^2 B_{, i} d t d x^i & \\
    +a^2\left[(1-2 \Phi) \delta_{i j}+2 E_{, i j}\right] d x^i d x^j,
\end{aligned}    
    \label{eq:SVTmetric}
\end{equation}
where the four scalar functions $\Phi, \Psi, B$ and $E$ define the $3+1$ perturbations of the full ADM metric and $a$ the usual background scale factor. The full dictionary from ADM to the $1^{\rm st}$ order scalar decomposition follows from the identification of the metrics
\begin{equation}
\left \{
    \begin{array}{ll}
        \alpha & =\alpha_b(1+\Psi), \\
         \beta_i &=-a^2 B_{, i}, \\
         \gamma_{i j}& =a^2\left[(1-2 \Phi) \delta_{i j}+2 E_{, i j}\right],\\
    \end{array} \right .
        \label{eq:dictionnarySVT}
\end{equation}
which leads to the linear extrinsic curvature
\begin{equation}
\left \{
    \begin{aligned}
 K^i{}_j & =-H \delta_j^i+\frac{1}{3} \kappa \delta_j^i -\left(\partial^i \partial_j-\frac{1}{3} \Delta \delta_j^i\right) \chi, \\
        \chi & \equiv-\frac{a^2}{\alpha_b}(B-\dot{E}), \\
        \kappa & \equiv 3\left(\frac{\dot{\Phi}}{\alpha_b}+H \Psi\right)-\Delta \chi,\\
    \end{aligned} \right .
        \label{eq:dictionnarySVT2}
\end{equation}
where $\Delta=\partial_i \partial^i=\nabla^2 / a^2$ and the energy-momentum densities
\begin{equation}
\left \{
\begin{aligned}
        \rho & = \rho_b+\frac{\dot{\phi_b}}{\alpha_b}\frac{\delta\dot{\phi}}{\alpha_b}+\frac{dV}{d\phi}(\phi_b)\delta\phi-\frac{{\dot{\phi_b}}^2}{\alpha_b^2}\Psi, \\
        \quad {\cal J}_i & =  -\frac{\dot{\phi_b}}{\alpha_b}\partial_i\delta \phi,\\
        S_{i j} & = (P_b+\delta P) \delta_{i j}+\left(\partial_i \partial_j-\frac{1}{3} \delta_{i j} \nabla^2\right) \sigma,\\
\end{aligned}\right .
\end{equation}
 where $\sigma = 0$ (no anisotropic shear in this work, which is gauge-invariant to say), which means that $\tilde{S}_{i j} =0$ at first order. Note that the 3-Ricci tensor is also very simple and is given by 
 \begin{equation}
     { }^{3} R^i{}_j =\Delta \Phi \delta^i{}_j+\Phi^{,i}{ }_{,j},
 \end{equation} 
at first order. 

Substituting these expressions into the full ADM equations yields a set of $0^{\rm th}$ and $1^{\rm st}$ order perturbation equations which do not yet incorporate a choice of gauge. Setting $H \equiv -\frac{1}{3} K_b = \frac{1}{\alpha_b}\frac{\dot{a}}{a} $, the background terms yield
\begin{equation}
\left\{\begin{aligned}
6 H^2 - 2M_{Pl}^{-2}\left(V(\phi_b)+\frac{1}{2}\frac{\dot{\phi_b}^2}{\alpha_b^2}\right) & =  0, \\
3M_{Pl}^{-2}V(\phi_b)-9 H^2 - 3\frac{\dot{H}}{\alpha_b} & = 0,\\
\frac{\ddot{\phi_b}}{\alpha_b^2}+\left(3H-\frac{\dot{\alpha_b}}{\alpha_b^2}\right)\frac{\dot{\phi_b}}{\alpha_b}+ \frac{dV}{d\phi}(\phi_b)  & = 0,
\end{aligned}\right .
    \label{eq:backgroundEq}
\end{equation}
from the Hamiltonian constraint, the extrinsic curvature, and the field evolution equations respectively, while other equations are vanishing. These are of course the equations of homogeneous and isotropic FLRW cosmology. At first order, the gravitational evolution equations become:
\begin{equation}
\left\{\begin{aligned}
        \frac{\dot{\kappa}}{\alpha_b}+2 H \kappa+\left(\Delta+3 \frac{\dot{H}}{ \alpha_b} \right)\Psi -\frac{1}{2}M_{Pl}^{-2}(\delta \rho+3 \delta P) = 0, &\\
         \left(\frac{1}{3} \delta_j^i \nabla^2-\partial_i \partial_j\right)\left(\frac{\dot{\chi}}{ \alpha_b}+H \chi-\Psi+\Phi- M_{Pl}^{-2} \sigma\right)=0, &
\end{aligned}\right .
    \label{eq:Kcpt_eq}
\end{equation}
while for the field equation 
\begin{equation}
\begin{aligned}
    \frac{\delta\ddot{\phi}}{\alpha_b^2}+\left(3 H-\frac{\dot{\alpha_b}} { \alpha_b^2}\right) \frac{\delta \dot{\phi}}{\alpha_b}-\Delta \delta \phi+\frac{d^2 V}{d \phi^2}(\phi_b) \delta \phi& \\
    -\frac{\dot{\phi_b}}{\alpha_b}\left(\kappa+\frac{\dot{\Psi}} { \alpha_b}-3 H \Psi\right)+2 \Psi \frac{d V}{d \phi}(\phi_b)=0. 
    \end{aligned}
        \label{eq:fieldCPT}
\end{equation}
Finally, the first-order constraint equations are:
\begin{equation}
\left \{
\begin{aligned}
        4\Delta \Phi-4H \kappa-2M_{Pl}^{-2}\delta \rho = 0, &\\
         -2\partial_i\left[\Delta \chi+\kappa+-\frac{3}{2}M_{Pl}^{-2}\frac{\dot{\phi_b}}{\alpha_b}\delta \phi \right]= 0. &
\end{aligned}\right .
    \label{eq:ConstraintsCPT}
\end{equation}
Hence we have five equations for the five unknowns $\delta \phi, \Psi, \Phi, E$ and $B$.  In general, we can use the gauge freedom to fix two of these variables. The two constraints then further reduce the dynamical scalar degrees of freedom down to one, usually encapsulated in a variable that is also gauge invariant.

\section{Spacetime foliations and scalar fields from the comoving curvature perturbation \label{sec:secIII}}

In this section, we will explicitly show how a choice of gauge combined with knowledge of the gauge invariant variable $\mathcal{R}$ determine, to first order, the 3D spacetime hypersurfaces corresponding to the ADM foliation and all perturbation variables. We will use the relations of this section later when we derive our stochastic equations. We start by recalling a set of well-known gauge invariant variables used in the literature.

In the following and unless stated differently, $\alpha_b = 1$, which implies that $\dot{()}$ is a cosmological time derivative and $H$ is the common Hubble rate.

\subsection{Scalar perturbations, all-in-one  \label{sec:secIIIA}}

Gauge-invariant quantities present many advantages beyond their invariance under coordinates transformations: they naturally reduce the dynamics to the single dynamical scalar degree of freedom and their dynamical equations guarantee that the constraints of General Relativity are automatically satisfied. Many linear combinations of $\delta \phi, \Psi, E, B$ and $\Phi$ are known to be gauge-invariant at first order, in particular, Bardeen potentials $\Phi_B$ and $\Psi_B$ \cite{Bardeen1980}, the gauge-invariant field perturbation $\delta\phi_{gi}$, the curvature perturbation on uniform density hypersurfaces $\zeta_{gi}$ or the curvature perturbation on comoving hypersurfaces $\mathcal{R}$. They are given by
\begin{equation}
    \left \{\begin{array}{cl}
         \Phi_B & = \Phi + H\chi, \\
         \Psi_B & = \Psi - \dot{\chi},\\
         \delta\phi_{gi} & = \delta\phi-{\dot{\phi_b}}\chi,\\
         \zeta_{gi} &= -\Phi+ \frac{\delta\rho}{3(\rho_b+ P_b)}, \\
         {\cal R} & = \Phi + \frac{H}{\dot{\phi_b}}\delta \phi.
    \end{array}\right .
    \label{eq:GIquantities}
\end{equation}
Those are convenient for the community as they appear to have simple gauge-invariant evolution equations. For example, for a single field $[P(X,\phi) = X-V(\phi)]$-theory, the Fourier modes of ${\cal R}$ satisfy \cite{Mukhanov1981,brandenberger1993classical}
\begin{equation}
\ddot{{\cal R}}_k + H(3-\varepsilon_2)\dot{{\cal R}}_k+ \frac{k^2}{a^2}{\cal R}_k = 0,
    \label{eq:R_evol_eq}
\end{equation}
where $\varepsilon_2 = -\dot{\varepsilon_1}/H\varepsilon_1$ and $\varepsilon_1 = -\dot{H}/H^2$ are the slow-roll parameters. A remarkable feature of the above equation is the perfect cancellation of ${\cal R}$'s effective mass term.
As we mentioned above, by knowing the background and a universal description of the perturbations such as ${\cal R}$, one can then write all fields in any gauge, including ones employed in Numerical Relativity. Examples are given in the next subsections.

In this work, we will use $\mathcal{R}$ as our master gauge-invariant variable. It is possible to express all other gauge invariant variables in terms of $\mathcal{R}$ as 
\begin{equation}
    \left\{\begin{array}{cl}
         \Phi_{B}  & = -\varepsilon_1 H a^2 k^{-2}\dot{{\cal R}}, \\
         \Psi_{B}  & = \varepsilon_1{\cal R}  + \varepsilon_1a^2k^{-2}\left[\ddot{{\cal R}} +H(2-\varepsilon_2)\dot{{\cal R}} \right],\\
         \delta\phi_{gi}  & =  \sqrt{2\varepsilon_1}M_{Pl}\left[{\cal R} +\varepsilon_1a^2 H k^{-2}\dot{{\cal R}} \right],\\
         \zeta_{gi}  &= -{\cal R} +\frac{1}{3} H^{-1}\dot{{\cal R}}. \\
    \end{array}\right .
    \label{eq:GaugeInvExplicit}
\end{equation}
These relations are most easily obtained in the comoving gauge (see below) but since they are relations between gauge invariant variables they hold in \emph{all} gauges. Eqn.~(\ref{eq:GaugeInvExplicit}) will be more than useful in some of the gauges we will examine below when the constraint equations will not be directly solvable. Note also that by using eqn.~(\ref{eq:R_evol_eq}) one easily shows that $\Phi_B=\Psi_B$ which is known to hold for a scalar field as there is no anisotropic stress at linear order. We will however keep these two variables distinct as they will differ by a stochastic term when we consider coarse-graining - see eqn.~(\ref{eq:GaugeInvExplicit>}).

\subsection{Spacetime foliations in different gauges \label{sec:secIIIB}}
\subsubsection{Pure gravity - comoving gauge \label{sec:secIIIB1}} A gauge will be called \textit{pure gravity} if it sets $\delta\phi$ and $E$ to small arbitrary spacetime functions $\delta\phi^*$ and $E^*$, their smallness being of the order of other perturbative quantities such as ${\cal R} $. The null case $\delta\phi^*=E^*=0$ is commonly used in cosmology and is referred to as the \textit{comoving} gauge. Such a pure gravity gauge is probably among the simplest ones to deduce the whole 3D hypersurface. Indeed, $\Phi $ is given directly via the definition of ${\cal R} $ together with the chosen $\delta\phi^*$. Substituting it in the linearized energy and momentum constraints (eqn.(\ref{eq:ConstraintsCPT}) yields the result for $\Psi $ and $B $
\begin{equation}
\left \{
\begin{array}{cl}
 E   & = E^*,\\
     \delta\phi    & = \delta\phi^*,\\
    \Phi    & = {\cal R}  -(2 \varepsilon_1M_{Pl}^2)^{-1/2}\delta\phi^*, \\
    \Psi  & = (2 \varepsilon_1M_{Pl}^2)^{-1/2}(\varepsilon_1+\frac{1}{2}\varepsilon_2)\delta\phi^*\\
    & -H^{-1}\dot{{\cal R}}  + (2 \varepsilon_1M_{Pl}^2H^2)^{-1/2}\dot{\delta\phi^*},\\
    B  & = \dot{E}^*+k^{-2}\varepsilon_1\dot{\cal R}  \\& +(a^2H)^{-1}\left[{\cal R}  -(2 \varepsilon_1M_{Pl}^2)^{-1/2}\delta\phi^*\right].\\
\end{array}\right .
    \label{eq:PureGravGaugeDecomp}
\end{equation}
Alternatively, we could have used eqn.~(\ref{eq:GaugeInvExplicit}) directly to obtain the same relations. Either way, all quantities have been expressed in terms of ${\cal R} $, and the constraints are satisfied by construction.

For the comoving gauge we have, in particular
\begin{equation}
\left \{
\begin{array}{cl}
 E   & = 0,\\
     \delta\phi    & = 0,\\
    \Phi    & = {\cal R}, \\
    \Psi  & = -H^{-1}\dot{{\cal R}},  \\
    B  & = k^{-2}\varepsilon_1\dot{\cal R}  +(a^2H)^{-1}{\cal R},  \\
\end{array}\right .
    \label{eq:ComGauge}
\end{equation}
and both the foliation and the spatial geometry are expressed in terms of ${\cal R}$.

\subsubsection{Fixed spatial curvature - spatially flat gauge\label{sec:secIIIB2}} A gauge will have \textit{fixed spatial curvature} if it sets $\Phi$ and $E$ to small arbitrary spacetime functions $\Phi^*$ and $E^*$. The null case $\Phi^*=E^*=0$ corresponds to the well-known spatially flat gauge of cosmological perturbation theory.

This family of gauges is as easy to handle as the pure gravity family above. Noting that ${\cal R} $ is proportional to one of the perturbation quantities, here $\delta\phi$, the Momentum constraint can be used to obtain $\Psi$ while the Hamiltonian constraint gives $B$ 
\begin{equation}
\left \{
\begin{array}{cl}
 E   & = E^*,\\
  \Phi  & = \Phi^*,\\
    \delta\phi    & = \sqrt{2\varepsilon_1}M_{Pl}({\cal R} -\Phi^*),\\
    \Psi    & =\varepsilon_1({\cal R} -\Phi^*)-H^{-1} \dot{\Phi}^*,\\
   
    B  & = \dot{E}^*+\varepsilon_1k^{-2}\dot{\cal R} +(Ha^2)^{-1}\Phi^*.\\
\end{array}\right .
    \label{eq:FixedRGaugeDecomp}
\end{equation}
Again, we could have used eqn.~(\ref{eq:GaugeInvExplicit}) directly to obtain the same relations. We have therefore obtained the foliation and the field perturbation in terms of the only degree of freedom for the evolution of the perturbations. The hypersurface geometry is of course explicitly given by the gauge choice here.

\subsubsection{Newtonian gauge \label{sec:secIIIB3}} A gauge will be called \textit{Newtonian} if it sets $B$ and $E$ to small arbitrary spacetime functions $B^*$ and $E^*$. Unlike the two previous gauge families, in this type of gauge using the the constraints alone is not enough to express all quantities in terms of $\mathcal{R}$. We can however resort to eqn.~(\ref{eq:GaugeInvExplicit}). Starting from $ \chi^*  = -a^2(B^*-\dot{E}^*)$, one gets $\Phi $ from $\Phi_B $ and then $\Psi $ is obtained from either the Momentum constraint or $\Psi_B $
\begin{equation}
\left \{
\begin{array}{cl}
     E   & = E^*, \\
     B  & = B^*, \\
    \Phi    & = \Phi_{B}  -H \chi^* =   -\varepsilon_1 Ha^2 k^{-2}\dot{{\cal R}}-H \chi^*, \\
    \delta\phi    & = \sqrt{2\varepsilon_1 }M_{Pl}\left[{\cal R}  +  \frac{\varepsilon_1 Ha^2}{k^2}\dot{{\cal R}} -a^2(B^*-\dot{E}^*) \right],\\
    \Psi  & = -\varepsilon_1 a^2 H k^{-2} \dot{\cal R},  \\
    & -2a^2H(B^*-\dot{E}^*)-a^2(\dot{B}^*-\ddot{E}^*).
\end{array}\right .
    \label{eq:NewtGaugeDecomp}
\end{equation}
The standard case in cosmology is of course obtained for $B^*=E^*=0$.

\subsubsection{Uniform ${\cal N}$ \label{sec:secIIIB4}} 
If one defines $\cal{N}$ to be the number of e-folds elapsed from an arbitrary time $t_0$ via
\begin{equation}
\left\{
\begin{aligned}
    \partial_0{\cal N} \equiv -\frac{1}{3}\alpha K =  \frac{1}{6} \partial_0 \ln {^{(3)}g},\\
    {\cal N}(t_0,\vec{x}) = 0,
\end{aligned}\right .
\end{equation}
then a gauge will be of \textit{uniform} (or \emph{fixed}) ${\cal N}$ if $\delta{\cal N}$ is fixed to a small arbitrary spacetime function $\delta{\cal N}^*$ while $B = 0$. As explained in sec. \ref{sec:secIIC2}, this type of gauge with $\delta{\cal N}^* = 0$ is the one mainly used in the formulations of Stochastic Inflation in the current literature.

As ${\cal N} \equiv -\frac{1}{3}\int_{t_0} K \alpha dt $, one gets ${\cal N}_b = \int_{t_0} H dt$ at zeroth order and  $\delta{\cal N} = \Phi+\frac{1}{3}k^2E$ at first order \cite{pattison_stochastic_2019}. To calculate the whole hypersurface, one first needs to solve for $E $ by using $\Phi_{B }$ to obtain
\begin{equation}
    \dot{E}  -\frac{1}{3}a^{-2}H^{-1}k^2 E  = -a^{-2}H^{-1}\delta{\cal N}^* - \varepsilon_1 k^{-2} \dot{\cal R}, 
    \label{eq:prepUniNGauge}
\end{equation}
the solution of which provides $E$ as
\begin{equation}
    E= e^{\int^{t}_{t_0}\!\frac{k^2}{3a^2H}}\left[E_0(\vec{x})-\int^t_{t_0}e^{-\int^{s}_{t_0}\!\frac{k^2}{3a^2H}dt'}\left( \frac{\delta{\cal N}^*}{a^2H} +\frac{\varepsilon_1  \dot{\cal R}}{k^2}  \right)ds\right],
\end{equation}
with $E(t_0,\vec{x})\equiv E_0(\vec{x})$. Note that there is still some gauge freedom left in the choice of the initial value of $E$ in space. Once $E $ is known, one gets
\begin{equation}
\left \{
\begin{array}{cl}
  \Phi  & = -\frac{1}{3}k^2E +\delta{\cal N}^*,\\
    \delta\phi    & = \sqrt{2\varepsilon_1}M_{Pl}({\cal R} +\frac{1}{3}k^2E -\delta{\cal N}^*),\\
    \Psi    & =\varepsilon_1({\cal R} +\frac{1}{3}k^2E -\delta{\cal N}^*)+H^{-1} (\frac{1}{3}k^2 \dot{E} -\dot{\delta{\cal N}}^*).\\
\end{array}\right .
    \label{eq:UniNGaugeDecomp}
\end{equation}

\subsubsection{Generalised synchronous gauges \label{sec:secIIIB5}}
A gauge will be called \textit{generalised synchronous} if it sets $\Psi$ and $B$ to small arbitrary spacetime functions $\Psi^*$ and $B^*$ to stay within perturbation theory. The null case corresponds to the synchronous gauge in cosmology. Similarly to the Newtonian case, in this gauge family constraints are not enough to solve in terms of ${\cal R} $, which is why we use eqn.(\ref{eq:GaugeInvExplicit}) on top of it. This inconvenience is an indication of the gap between numerical relativity and cosmology gauges.

Using the definitions of the gauge-invariant $\Psi_B $ and $\Phi_B$ and the latter's relation to $\cal{R}$, see (\ref{eq:GaugeInvExplicit}), we can obtain the following cascade
\begin{equation}
\left \{
\begin{array}{cl}
 \chi    & = \int^t (\Psi^*  -\Psi_B  ) dt' + \chi_0,\\
     E   & = \int^t [B^*  +a^{-2}\chi   ]dt' +E_0,   \\
    \Phi    & = \Phi_{B}  -H \chi,    \\
    \delta\phi    & = \sqrt{2\varepsilon_1 }M_{Pl}[{\cal R}   - \Phi   ],\\
\end{array}\right .
    \label{eq:ADMgaugeDecomp}
\end{equation}
which are all functions of ${\cal R} $ and its $1^{st}$ and $2^{nd}$ order time derivatives and their integrals.

In this gauge both $\chi $ and $E $ need to be initialized at a given time by space-only functions $\chi_0(\vec{x})$ and $E_0(\vec{x})$. It is indeed a well-known problem that the synchronous gauge (and so more generally any small extended synchronous gauge) does not fix all gauge degrees of freedom and that there is still spatial dependence \cite{baumgarte_shapiro_2010}, even though this can sometimes be confused with physical choices in the literature, see Appendix \ref{app:appB} for proof. In this work, we will choose gauges such that $\chi_0=E_0=0$  at an arbitrary time, hidden in an implicit boundary of the time integrals.

\section{Stochastic equations for General Relativity \label{sec:secIV}}

\subsection{Schematic formulation of effective stochastic IR dynamics  \label{sec:secIIC}}

As mentioned in the introduction, Stochastic Inflation can be thought of as an Effective Field Theory (EFT) \cite{weinberg_development_2021} of QFT on curved spacetime (QFTCS), valid on scales where quantum correlation functions can be well approximated by classical, stochastic ones. Its utility lies in treating fluctuations beyond perturbation theory and associated truncations in orders of non-linearity on super-Hubble scales, compared to QFTCS \cite{cable_second-order_2022}. 

By definition, the EFT refers to variables that are coarse-grained beyond a certain length scale, normally commensurate with the Hubble radius, but there is no general method for this coarse-graining in inflationary spacetimes. The greatest difficulty of applying known EFT techniques with SI is probably the nature of its IR-UV split: the cutoff is spacetime-dependent. While building an EFT from a path integral \cite{hosoya_stochastic_1989,prokopec_path_2010} is a possible approach in simple cases of test fields in de Sitter (dS), no such approach that also includes the fluctuations of the metric in inflationary spacetimes has been achieved.

In this work, we use a coarse-graining of the equations of motion (EOM), which is probably the most common approach for reasons that will become clearer later. Given the complexity of the complete set of equations, we first schematically review its philosophy. Let's assume that we possess an IR classical theory giving access to some second-order tensorial partial differential equations and their linearization at first order in CPT of the form
\begin{equation}
\left\{
\begin{array}{ll}
    \Lambda^{iab}\nabla_a\nabla_b X^i + \Omega^i & = 0,\\
    \lambda^{iab}\partial_a\partial_b \delta X^i + \delta\Omega^i & = 0,\\
\end{array}\right .
    \label{eq:GeneralSIstart}
\end{equation}
where $X^i$, $\Omega^i(X)$, are tensors of identical rank with $\Omega^i$ being functions of the $X^i$. $\delta X^i$ and $\delta\Omega^i$ are their linearized perturbations, $\delta\Omega^i$ being a linear combination of $\delta X^i$ and their partial derivatives.  Those equations could be the previous section's or approximated versions of them or any others. No assumption is made on $\Lambda$, $\Omega$, $\lambda$, and $\delta\Omega$ except their smoothness and that they depend on $\{X^j,\nabla X^j\}_j$, and spacetime coordinates. No straightforward link between $\lambda$ and $\Lambda$ or $\delta\Omega$ and $\Omega$ can be written down, although one has to bear in mind that $\delta\Omega^i$ can carry perturbations of the $\{X^j\}_j$ and of $g$ along with their first partial derivatives. Note that the background equation has been kept implicit but answers the relation $X^i = X_b^i+\delta X^i.$  The background is most of the time assumed homogeneous, which will be the case in this work.

The next step is to find a solution of the first-order equations in Fourier space for $\{\delta X^j_{\vec{k}}\}_j$, in at least one gauge. In the case of Inflation, the spatially flat gauge is the most common one to use where the dynamical field is equal to the curvature perturbation on comoving hypersurfaces following the Mukhanov-Sasaki equation, eqn.(\ref{eq:R_evol_eq}). After that, a gauge transform can bring the calculated spectrum to the desired gauge, although those extra terms are usually neglected under the right assumptions \cite{pattison_stochastic_2019}. The $\{\delta X^j\}_j$ are assumed to remain dynamical quantities in this new gauge, i.e. with a non-zero conjugate momentum in the sense of Field Theory (FT). This assumption is necessary to make sure the equations are dynamical but also to make sure that canonical quantization is possible. At this stage, quantization is usually written down explicitly as an expansion in the modes
 \begin{equation}
     \delta\hat{X}^i = {\cal F}^{-1}\{ \delta X^i_{\vec{k}}\hat{a}_{\vec{k}}\} + h.c,
         \label{eq:ModesExpGeneral}
 \end{equation}
 where $[\hat{a}^{}_{\vec{k}},\hat{a}_{\vec{k}'}^\dagger ] =  \delta^{(3)}(\vec{k}-\vec{k}')$ and any other commutator being $0$. This thus gives us a Gaussian spectrum.

With the solutions of the linearized equations at hand, the principle of the literature is simple and is called IR-UV splitting \cite{Starobinsky_stochastic_1988}. The quantised perturbations are indeed split as $\delta \hat{X}^i = \delta \hat{X}^{i>}+\delta \hat{X}^{i<}$, i.e. into long and short wavelength parts respectively\footnote{Note that many authors use $>$ and $<$ for long and short wavenumbers instead.}, for instance with a spacetime-dependent window function in Fourier space
 \begin{equation}
 \delta \hat{X}^{i>} = {\cal F}^{-1}\{ W^i_{\vec k}\delta X^i_{\vec{k}}\hat{a}_{\vec{k}}\} + h.c,
     \label{eq:window}
 \end{equation}
For SI, the window function would typically let short modes become long ones when their wavelegth equals or exceeds the Hubble horizon, namely $k\lesssim aH$, to ensure classicalisation.
 
 The UV-IR splitting via some window function implies that the linearized equation for the long wavelength $\delta X^{i>}$, simplified by the use of $X_b^i$'s and $\delta X^i$'s equations, will have a \emph{non-zero right-hand side (RHS) term} (compare with eq.~(\ref{eq:GeneralSIstart}))
\begin{equation}
    \lambda^{iab}\partial_a\partial_b \delta \hat{X}^{i>} + \delta\hat{\Omega}^{i>} = \hat{\Sigma}^i,
        \label{eq:GeneralCOarseGrain}
\end{equation}
where the RHS $\hat{\Sigma}^i$ is a Fourier expansion of functions of $\delta \hat{X}^{i>}$ and its first derivatives, but also of the window function derivatives.\footnote{Some authors \cite{rigopoulos_non-linear_2005} will prefer to calculate the LHS of eqn.(\ref{eq:GeneralCOarseGrain}) given a windowed $\delta X^{i>}$ while others \cite{Starobinsky_stochastic_1988,pattison_stochastic_2019} would equivalently calculate $\hat{\Sigma} = -\lambda^{iab}\partial_a\partial_b \delta \hat{X}^i_< - \delta\hat{\Omega}^i_<$ given a windowed $\delta X_< ^i$.} Eq.~(\ref{eq:GeneralCOarseGrain}) is a quantum Langevin equation \cite{nakao_stochastic_1988} obtained after coarse-graining and quantisation. To get a classical equation, i.e. to get rid of the operators, we need the Stochastic approximation to write
 \begin{equation}
     \delta\hat{X}^{i>} \simeq \frac{1}{\sqrt{2}}[{\cal F}^{-1}\{ \delta X^i_{\vec{k}}{}^>\boldsymbol{\alpha_{\vec{k}}}\} + c.c],
         \label{eq:GeneralStochasticApprox}
 \end{equation}
where $\boldsymbol{\alpha_{\vec{k}}}$ are now random variables such that
\begin{equation}
\left\{
\begin{aligned}
        \langle \boldsymbol{\alpha_{\vec{k}}}\boldsymbol{\alpha_{\vec{k}'}} \rangle_{\mathbb{P}}  & = 0, \\
           \langle \boldsymbol{\alpha_{\vec{k}}}\boldsymbol{\alpha_{\vec{k}'}}^* \rangle_{\mathbb{P}} & =  \delta^{(3)}(\vec{k}-\vec{k}').
\end{aligned}\right .
\end{equation}
To match the Gaussian spectrum of the linearized quantum operators, one should take $\alpha_{\vec{k}}\sim\mathbb{C}{\cal N}(0,1)$, which is equivalent to $Real[\alpha_{\vec{k}}],Im[\alpha_{\vec{k}}]\sim {\cal N} (0,1/2)$ independently or $|\alpha_{\vec{k}}|\sim Rayleigh(1/\sqrt{2})$ and $Arg[\alpha_{\vec{k}}]\sim {\cal U} (0,2\pi)$. The validity of this classicalisation can be tested by showing that correlators of $\delta \hat{X}^>$ and associated conjugate momenta receive negligible contributions from non-commutative terms on long wavelengths. For SI, the validity has already been studied and the following analysis will be restricted to the applicable cases \cite{Starobinsky_stochastic_1988, Polarski_1996,KIEFER_1998,burgess_minimal_2022}. 

Finally, under the assumption that linear theory is enough at horizon-crossing for computing the stochastic $\boldsymbol{\Sigma}^i$, the LHS of (\ref{eq:GeneralCOarseGrain}) can be promoted to the full non-linear equation in (\ref{eq:GeneralSIstart}) with the understanding that it refers to the coarse-grained quantity $X^{i>}$:
\begin{equation}
\Lambda^{iab}\nabla_a\nabla_b X^{i>} + \Omega^{i>} = \boldsymbol{\Sigma}^i.
    \label{eq:GeneralSI}
\end{equation}
The above equation constitutes the \emph{long wavelength, stochastic} version of the original and is postulated to furnish an adequate approximation to the long wavelength fluctuation dynamics. The line of reasoning leading to (\ref{eq:GeneralSI}), applied to a truncated subset of the Einstein equations, underlies most existing expositions of stochastic inflation equations in the literature.      

This linear source approximation, together with its assumed Gaussianity, is a common feature of the stochastic inflation literature \cite{Starobinsky_stochastic_1988, hosoya_stochastic_1989, grain_stochastic_2017}, but its justification is rarely made explicit. The validity domain of this approximation lies in scenarios where, before horizon-crossing, UV scales have suppressed higher-order statistics compared to the UV tree-level or compared to their IR counterparts generated from the UV tree-level, after horizon-crossing. A quasi-dS universe is probably the safest case in this regard, which is convenient for its believed physical relevance. Some previous work \cite{cohen_stochastic_2021} has even managed to account for suppressed next-order statistics in the noise. More generally, this stochastic evolution should be reliable for studying any non-linearity and non-gaussianity generated by the evolution that is larger than the first order in CPT. However, if no non-perturbative effect is obtained from SI, then one should stick to QFTCS, valid on IR scales and beyond tree-level before horizon-crossing, which is more precise than SI, as the disagreement between the two shows in perturbative regimes \cite{cruces_stochastic_2022}.

In this section and following the literature, the stochastic backreaction (different from the classical Einstein backreaction) has also been neglected because we have linearised the theory around the homogeneous background and not around the IR quantities by invoking the Starobinsky approximation $X^{i>} = X_b^i+ O(\delta X ^i )$ \cite{Starobinsky_stochastic_1988}.  This includes the calculus of the spectrum of $\{\delta X^j_{\vec{k}}\}_j$.
This assumption is crucial to get closer to a Markovian system and at least an additive noise, thus facilitating analytical solutions. 

However, different heuristics have been used in the past for the stochastic update of those RHS and of the background quantities on which the modes evolve. Usually, background quantities $X_b^i$ in the RHS of eqn. (\ref{eq:GeneralSI}) and in the equation of $\{\delta X^j_{\vec{k}}\}_j$ are taken as their local IR versions $ X^{i>}(t,\vec{x})$. In this case, solving the equations requires a numerical approach \cite{levasseur_backreaction_2015,Figueroa22,cruces_stochastic_2022}. These studies have all reported a very small impact so far. In this work, we will keep it arbitrary and keep any RHS amplitude factor such as $H$ or $\varepsilon_i$ undefined in this respect, until further comment. In the linear limit, one always retrieves eqn. (\ref{eq:GeneralCOarseGrain}).

To summarize, stochastic inflation is about reducing QFTCS approximations to UV scales only, allowing for the possibility to study fully non-linear and non-perturbative phenomena on IR scales, i.e. above Hubble scales in our case. The Langevin-type EOM of stochastic inflation can also be mapped to an associated Fokker-Planck (FP) equation and both can be solved either analytically in some simple cases or numerically. From these solutions, one can then derive non-perturbative correlators or even the full probability functional of the fields. It is in this ability to provide results beyond perturbation theory or even fixed $n$-point correlators where Stochastic inflation's appeal lies.

\subsection{Current approaches to Stochastic Inflationary dynamics\label{sec:secIIC2}}

So far we presented a schematic picture of how one might obtain equations describing the dynamics of Stochastic Inflation from coarse-graining the equations of motion. However, applying the above scheme to the full equations of General Relativity has not been implemented due to the latter's relative complexity, the a priori large number of variables, which can be both dynamical and constrained, and the necessity of making coordinate/gauge choices. As a result, various approximations have been made, mostly taking the long wavelength limit and reducing the number of dynamical variables.  

Early work focused on the coarse-graining of the Klein-Gordon field equation only, assuming an unperturbed (initially dS) background and coarse-graining the field and sometimes its conjugate momentum \cite{Starobinsky_stochastic_1988,sasaki_classical_1988,nambu_stochastic_1988,nakao_stochastic_1988,morikawa_dissipation_1990}. It was only in \cite{salopek_nonlinear_1990,salopek_stochastic_1991} that the first stochastic equations were formulated fully including metric perturbations and so backreaction. This approach is still widely used. It consists in decreasing the interdependence of GR equations by using the long wavelength approximation and judicious gauge choices. In the full ADM formalism, even before linearization and coarse-graining, the lowest order gradient expansion of eqn.\ (\ref{eq:Keq}) and (\ref{eq:cons_eq}) in the convenient $\beta^i = 0$ slicing becomes \cite{salopek_nonlinear_1990,rigopoulos_separate_2003,prokopec_ensuremathdeltan_2021} 
\begin{equation}
\left \{
\begin{aligned}
 \dot{K}-\alpha K^2
 - M_{Pl}^{-2} \alpha\left(\frac{1}{2} S-\frac{3}{2} \rho\right) = 0, & \\
  \dot{\tilde{K}}^i{}_j-\alpha K \tilde{K}^i{}_j + M_{Pl}^{-2} \alpha \tilde{S}^i{}_j=0, &\\
    \frac{1}{\alpha}\dot{\Pi}-K \Pi+\frac{d V}{d \phi}=0, & \\
\end{aligned}\right .
    \label{eq:longGR}
\end{equation}
together with the constraints
\begin{equation}
\left \{
\begin{aligned}
 \frac{2}{3} K^2-\tilde{K}_{i j} \tilde{K}^{i j}-2 M_{Pl}^{-2} \rho = 0, &\\
 \tilde{K}^j{}_{i \mid j}-\frac{2}{3} K_{\mid i}- M_{Pl}^{-2} {\cal { J}}_i=0. & 
\end{aligned}\right .
    \label{eq:longCons}
\end{equation}
Furthermore, $\tilde{S}^i{}_j$ is usually set to zero in the absence of anisotropic fluid sources, or because any possible contributions to it are considered higher order in the gradient expansion, leading to an exponential decay of ${\tilde{K}}^i{}_j$. This results in one equation less with only $K$ and $\Pi$ dynamics remaining, that is,
\begin{equation}
\left \{
\begin{aligned}
 \dot{K}-\alpha K^2
 - M_{Pl}^{-2} \alpha\left(\frac{1}{2} S-\frac{3}{2} \rho\right) = 0, & \\
    \frac{1}{\alpha}\dot{\Pi}-K \Pi+\frac{d V}{d \phi}=0, & \\
     \frac{2}{3} K^2-2 M_{Pl}^{-2} \rho = 0, &\\
-\frac{2}{3} K_{\mid i}- M_{Pl}^{-2} {\cal { J}}_i=0. & 
\end{aligned}\right .
    \label{eq:SUA_GR}
\end{equation}
These equations form what is called the  \textit{separate universe approach} (SUA), widely used as a basis for the formulation of stochastic inflation. 

In many later works stemming largely from e.g.~\cite{wands_new_2000,Lyth03}, the momentum constraint is not considered part of the long wavelength approximation, an approach recently referred to as \textit{$(k$$=$$0)$-SUA} in \cite{cruces_review_2022}. The relevant literature then interprets literally the similarity of the $\Pi$, $K$ and the Hamiltonian constraint in eqn.~(\ref{eq:SUA_GR}) with Friedman's background equations of cosmology eqn.~(\ref{eq:backgroundEq}). The local quantity $\frac{1}{3}K(x)$ replaces the homogeneous Hubble parameter $H$, so that the whole inhomogeneous universe is described as made up of patches, each of which evolves independently. The EOM coarse-graining has largely been applied to those two equations only: linearising provides the windowed RHS terms which can then be evaluated using UV modes solutions.

Unlike scalars on a fixed background, the inclusion of metric perturbations brings about the issue of the appropriate slicing to be used for both the linearisation and the noise calculation, especially beyond slow-roll scenarios. In this respect, the uniform-${\cal N}$ gauge slicing (see sec.\ \ref{sec:secIIIB4}) has emerged as the most natural, with \cite{Finelli09,Finelli10} arguing early on that in this slicing one recovers CPT equations and thus the long wavelength QFTCS limit when linearising the overdamped field equations. In general, the uniform-${\cal N}$ gauge has emerged as the more commonly used because it allows direct access to the statistics of the fields in terms of ${\cal N}_b$ slicing, a necessary step for the first-passage-time analysis (FTPA) and the associated stochastic $\Delta{\cal N}$ formalism which provides information about the non-linear curvature perturbation on uniform field hypersurfaces \cite{vennin_correlation_2015}. 

When performing the EOM coarse-graining in this gauge, one gets the SI equations \cite{pattison_stochastic_2019}, which we will refer to as the \textit{$(k$$=$$0)$-SUA}, adopting the term from \cite{cruces_review_2022}, or the `usual' equations
\begin{equation}
\left \{
\begin{aligned}
 &\frac{\partial \phi^>}{\partial{\cal N}_b} -\pi^> = \boldsymbol{\Sigma_{\phi}}, \\
 &\frac{\partial \pi^>}{\partial{\cal N}_b}   + (3-\varepsilon_1^>)\pi^>+\frac{1}{(H^>)^2}\frac{dV}{d\phi}(\phi^>)= \boldsymbol{\Sigma_{\pi}}, \\
& {H^>}^2  = \frac{V(\phi^>)}{1-\frac{1}{6M_{Pl}^2}(\pi^>)^2},
\end{aligned}\right .
    \label{eq:SIeq}
\end{equation}
where ${\cal N}_b$ is the background number of e-folds, $\varepsilon_1^>=-\partial_{{\cal N}_b}\ln H^>$ the first slow-roll IR parameter, and
\begin{equation}
\left \{
\begin{aligned}
\boldsymbol{\Sigma_{\phi}} & = +\frac{1}{\sqrt{2}}\left[{\cal F}^{-1}\left\{\frac{\partial W}{\partial{\cal N}_b}\delta \phi_{\vec{k}}\boldsymbol{\alpha_{\vec{k}}}\right\} + c.c\right], \\
\boldsymbol{\Sigma_{\pi}} & = +\frac{1}{\sqrt{2}}\left[{\cal F}^{-1}\left\{ \frac{\partial W}{\partial{\cal N}_b}\delta \pi_{\vec{k}}\boldsymbol{\alpha_{\vec{k}}}\right\} + c.c\right].
\end{aligned}\right .
    \label{eq:SInoise}
\end{equation}
In particular, this means that the coarse-graining has been assumed to apply as $\delta \phi ^> = W\delta \phi$ and $\delta \pi ^> = W\delta \pi$ in Fourier space \cite{pattison_stochastic_2019}.\footnote{Note that the authors write an opposite sign for the noise contributions when using $\delta X_< = W\delta X$ instead.} By defining $\pi^>$ with a stochastic source, the literature does not use the canonical momentum of $\phi^>$. One can also re-write the previous equations as a unique equation in the ADM $\Pi^> $,
\begin{equation}
\left \{
\begin{aligned}
\frac{\partial \phi^>}{\partial{\cal N}_b} &= \frac{{\Pi}^>} {H^>},  \\
 \frac{\partial {\Pi}^>}{\partial{\cal N}_b}  & +3{\Pi}^>+\frac{1}{H^>}\frac{dV}{d\phi}(\phi^>) = \boldsymbol{\Sigma}, \\
\end{aligned}\right .
    \label{eq:SIrewrite}
\end{equation}
with $ \boldsymbol{\Sigma} = H (\boldsymbol{\Sigma}_{\pi} + \partial_{{\cal N}_b}\boldsymbol{\Sigma}_{\phi}+(3-\varepsilon_1)\boldsymbol{\Sigma}_{\phi}) $, where $H$ is the background quantity or not depending on the scheme used, see sec.\ \ref{sec:secIIC}. In that sense, it is perfectly equivalent to coarse-grain the momentum and its definition (\textit{phase-space} coarse-graining), and to coarse-grain the momentum via the field only, which will be our approach, following the coarse-graining in Wilsonian EFTs \cite{moss_effective_2017}. It is common to keep the phase-space equations to apply the common overdamping assumption $|\delta \Pi|^2 \ll |\delta \phi|^2$ and thus reduce the problem to a first-order equation of the field only \cite{Starobinsky_stochastic_1988}. This assumption will not be made in the following. 

The role of the momentum constraint has been noted (refer e.g.\ to  \cite{salopek_nonlinear_1990,kodama_evolution_1998,rigopoulos_separate_2003,prokopec_ensuremathdeltan_2021,cruces_stochastic_2022,cruces_review_2022}) as being the only remaining link between the separate universes in the SUA. Neglecting it by invoking the long-wavelength approximation could lead to inconsistencies in CPT \cite{kodama_evolution_1998,Sasaki98,cruces_stochastic_2022}. This issue has been addressed in recent years by tackling the regime of validity of the \textit{$(k$$=$$0)$-SUA} by comparing to CPT \cite{pattison_stochastic_2019, artigas_hamiltonian_2022}: it seems that working far enough from the horizon-crossing scale with the uniform-${\cal N}$ gauge is a safe choice to match CPT. Nevertheless, the notable results achieved by this approach \cite{vennin_correlation_2015,pattison_quantum_2017,ezquiaga_exponential_2020} require transformations between gauges (flat gauge for the noise,  uniform-${\cal N}$ for the Langevin equations, and effectively uniform-$\phi$ in the FTPA) when giving up approximations such as slow-roll \cite{pattison_stochastic_2019}.

Other SI approaches have emerged in parallel with somewhat different assumptions. Starting from the Hamilton-Jacobi formalism developed in \cite{salopek_nonlinear_1990} from eqn.(\ref{eq:SUA_GR}), SI was formulated with the inclusion of the momentum constraint in \cite{salopek_stochastic_1991}, and the FPT approach of \cite{vennin_correlation_2015} was applied to this formalism in \cite{prokopec_ensuremathdeltan_2021, Rigopoulos:2021nhv}. Starting from the full set of (\ref{eq:SUA_GR}), a set of stochastic equations for non-linear variables like those given in (\ref{eq:NLgaugeinv}) were developed in 
\cite{rigopoulos_non-linear_2005} which by construction reduce to coarse-grained, gauge invariant CPT in all time-slicings with $\beta^i$$\,=\,$$0$. As shown in \cite{Rigopoulos:2005us, Tzavara:2010ge}, these equations produce the same perturbative results for the bispectrum as the $\Delta{\cal N}$. The more recent study \cite{cruces_stochastic_2022} also writes stochastic equations in a gauge other than uniform $\mathcal{N}$, the uniform Hubble gauge, by rehabilitating the momentum constraint in the gradient expansion. 

The mini-literature review discussed above attempts to provide a flavour of the current state of play regarding the status of SI formalism and the approximations that have been deemed necessary to develop it. In the next section, we demonstrate how one can do away with the long wavelength approximation and the SUA, providing a set of SI equations that retain both the scalar field and all the variables of the gravitational field, the only approximation being scalar linear CPT for the computation of the noise terms by coarse-graining. 

\subsection{Stochastic equations for ADM }

We now incorporate the continuous influx of modes that cross a smoothing scale commensurate with the Hubble scale, i.e.\ we apply the schematic procedure leading up to eqn.\ (\ref{eq:GeneralSI}) to the \emph{full set of Einstein equations}. This section contains the key results of this work, namely a computation of the stochastic sources associated with the ADM formulation of General Relativity. These stochastic equations are presented in (\ref{eq:CoarseGrainedADM}). We perform this computation in all the gauges discussed above, always finding the same result for the stochastic source terms, presented in eqns.~(\ref{eq:coarseGrainedReq}) and (\ref{eq:RHSADM}); as can be seen there, the source terms are always given by the same functional of the gauge invariant $\mathcal{R}$ and the chosen window function. The stochastic source terms appear on the RHS of the dynamical equations but not the constraint equations. We stress that we do not impose any gradient or slow-roll expansion.

\subsubsection{Coarse-graining linear theory\label{sec:secIVA} }
To coarse-grain GR we will apply the principle behind formula (\ref{eq:GeneralSI}) to determine the stochastic source terms from linear theory. As explained above, the major problem is choosing a gauge, its associated dynamical quantities, and the window. As we will see, in this section and the next one we perform a gauge-invariant coarse-graining, i.e.\ we coarse-grain the gauge-invariant linear theory encoded in the usual gauge invariant quantities of eqn.\ (\ref{eq:GIquantities}). In particular, the coarse-graining is made by using a time-dependent window function in Fourier space $W_k(t)$ on a gauge-invariant quantity, here ${\cal R}$. For the case of inflationary evolution, $W_k(t)$ would be activated after Hubble radius crossing when quantum modes can be assumed to behave classically. However, we stress that the coarse-graining method remains valid beyond this choice if one accepts operator-valued source terms that cannot be interpreted fully as classical random variables.

In practice, and as explained schematically through eqn. (\ref{eq:GeneralCOarseGrain}), coarse-graining means that we first search for the equations of the long-wavelength variable ${\cal R}_k^{>} =W_k {\cal R}_k $ where ${\cal R}$ obeys (\ref{eq:R_evol_eq}). ${\cal R}^{>}$ is still gauge-invariant of course but follows a slightly different equation of motion which can be obtained from eqn. (\ref{eq:R_evol_eq})
\begin{equation}
\ddot{{\cal R}}^{>}_k + H(3-\varepsilon_2)\dot{{\cal R}}^{>}_k+ \frac{k^2}{a^2}{\cal R}^{>}_k = {\cal S}_{{\cal R}},
    \label{eq:coarseGrainedReq}
\end{equation}
where the source term is 
\begin{equation}\label{eq:SR}
    {\cal S}_{{\cal R}}  = {\cal R}_k \ddot{W_k}+[2 \dot{\cal R}_k + (3-\varepsilon_2) H {\cal R}_k]\dot{W}_k.
\end{equation} 
\textit{We emphasize that in (\ref{eq:coarseGrainedReq}) the ${}^>$ operator has priority over time derivatives in all equations}, i.e. for any function ~$\dot{f}^> \equiv d\left(Wf\right)/dt$. Note also that the source term ${\cal S}_{{\cal R}}$ vanishes when the window is constant, i.e. usually when it is super-Hubble or sub-Hubble enough, the latter yielding  ${\cal R}^{>} \simeq 0$. This also shows that the existence of the source term is solely attributed to the time-dependent nature of the UV-IR split. 

Next, we derive the coarse-grained version of all the metric variables and the scalar field as given in sec. \ref{sec:secIIIB} by using the replacement 
\begin{equation}\label{eq:replacement}
{\cal R}_k\longrightarrow W_k {\cal R}_k,
\end{equation}
\emph{before taking any time derivative}. This is done for any of the gauge families discussed in that section. We stress that this prescription ensures that the constraint equations will stay perfectly satisfied at first order, which we have verified for all the gauges we examined.

\subsubsection{Coarse-graining General Relativity  \label{sec:secIVB}}

The expressions obtained after making the replacement (\ref{eq:replacement}) in the formulae of sec.~\ref{sec:secIIIB} can be substituted into the linearized evolution and constraint equations of section \ref{sec:secIIB} for any of the aforementioned gauges. As explained above, this procedure will provide the source terms for each of the ADM equations. Although, as we will see below, the final result is very simple and identical in all gauges, the full computation involves rather long expressions. We will therefore provide here a summary of the full derivation of the coarse-grained field equation in a small generalised synchronous for illustration. All the other stochastic ADM equations are obtained similarly.  

We need to calculate the $RHS$ term of the following linearized equation for the field perturbation:
\begin{equation}
\begin{aligned}
    \ddot{\delta\phi^>}+3 H \dot{\delta \phi^>}+\left(\frac{k^2}{a^2}+\frac{d^2V}{d\phi^2}(\phi_b)\right) \delta \phi^> +2 \Psi^* \frac{dV}{d\phi}(\phi_b)& \\
    -\dot{\phi_b}\left (3\dot{\Phi}^>+\dot{\Psi}^* +\frac{k^2}{a^2}\chi^>\right)
     = RHS({\cal R},W, \Psi^*, B^*),&
    \end{aligned}
\label{eq:LinearFieldCoarse}
\end{equation}
which is just the re-writing of eqn.~(\ref{eq:fieldCPT}), using the dictionary of eqns.~(\ref{eq:dictionnarySVT}), (\ref{eq:dictionnarySVT2}) in Fourier space. In this gauge, all relevant coarse-grained perturbation quantities are given in terms of $\mathcal{R}^>$ as
\begin{equation}
\left \{
\begin{array}{cl}
 \chi^>    & = \int^t [\Psi^*  -\Psi_B({\cal R}^>)  ] dt', \\
     E^>   & = \int^t [B^*  +a^{-2}\chi^>   ]dt',   \\
    \Phi^>    & = \Phi_{B}({\cal R}^>)  -H \chi^>  ,  \\
    \delta\phi^>    & = \sqrt{2\varepsilon_1 }M_{Pl}[{\cal R}^>   - \Phi^>   ],\\
\end{array}\right .
    \label{eq:ADMgaugeDecomp>}
\end{equation}
together with the coarse-grained Bardeen potentials
\begin{equation}
    \left\{\begin{array}{cl}
         \Phi_{B}({\cal R}^>)  & =-\varepsilon_1a^2k^{-2}H\dot{{\cal R}}^> ,\\
         \Psi_{B}({\cal R}^>)  & =  -\varepsilon_1a^2k^{-2}(H\dot{\cal R}^>-{\cal S}_{\cal R}).\\
    \end{array}\right .
    \label{eq:GaugeInvExplicit>}
\end{equation}
where the operator $>$ has priority over time derivatives.  This non-zero difference of the \emph{long wavelength} $\Phi_B^>$ and $\Psi_B^>$ is not due to anisotropic stress but simply due to the appearance of a time derivative in the definition of $\Psi_B$ in (\ref{eq:GIquantities}); when coarse-grained $\Psi_B$ acquires an extra stochastic source term compared to $\Phi_B$. It is a transient horizon-crossing effect.  This comes back to the usual equality for each mode if super-Hubble or deep sub-Hubble because the correction is negligible when the window function is constant.

Using the previous decomposition, further derivatives are needed
\begin{equation}
\left\{
\begin{aligned}
\dot{\delta \phi^>} & = \sqrt{2\varepsilon_1}M (H\Psi^* +\dot{\cal R}^>) - H(\varepsilon_1+\varepsilon_2/2)\delta\phi^>, \\
\dot{\Phi}^> & = -H\Psi^*+H\varepsilon_1{\cal R}^>-H\varepsilon_1\Phi^>,\\
\ddot{\delta \phi^>} & = \sqrt{2\varepsilon_1}M \left[{\cal S}_{\cal R}-(\varepsilon_1+\varepsilon_2/2)H\Psi^*-k^2a^{-2}{\cal R}^>+H\dot{\Psi}^* \right .\\
& \left .+H(\varepsilon_2/2-3)\dot{\cal R}^>\right] + H(3-\varepsilon_1-\varepsilon_2/2)\dot{\delta\phi^>}\\
& +H^2\left[H^{-2}V_{,\phi\phi}+\varepsilon_1(\varepsilon_1+\varepsilon_2/2)+\varepsilon_2(\varepsilon_1+\varepsilon_3/2) \right]\delta\phi^>,
    \end{aligned}\right .
        \label{eq:LinearFieldCoarseDecomp}
\end{equation}
where $V_{,\phi\phi}$ is the background quantity, function of $\varepsilon_1$, $\varepsilon_2$ and $\varepsilon_3$, see Appendix \ref{app:appA}. We now have all the coarse-grained variables needed to plug into the LHS of the coarse-grained field equation (\ref{eq:LinearFieldCoarse}). After a lengthy computation, a perfect cancellation occurs and only the ${\cal S}_{\cal R}$ term of $\ddot{\delta\phi^>}$ survives, giving
\begin{equation}
    RHS({\cal R},W, \Psi^*, B^*) = \sqrt{2\varepsilon_1}M_{Pl} {\cal S}_{\cal R},
        \label{eq:RHSLinearFieldCoarse}
\end{equation}
which is completely independent of $\Psi^*$ and $B^*$, i.e. of the specific functions corresponding to the choice of small gauge in which the computation was performed. Following the EOM approach we described in \ref{sec:secIIC}, we can now assume that stochastic source terms derived for the linear spectrum amplitude equations remain valid for their non-linear parent equation so that the final, non-linear super-Hubble Langevin equation in this gauge is
\begin{equation}
\begin{array}{r}
    \displaystyle\frac{1}{\alpha}\left(\dot{\Pi}+\beta^i \Pi_{\mid i}\right)-K \Pi-\frac{\alpha^{\mid i}}{\alpha} \phi_{\mid i} \\
 - \displaystyle\phi_{\mid i}^{\mid i}+ \displaystyle\frac{d V}{d \phi}=\sqrt{2\varepsilon_1}M_{Pl}{\cal F}^{-1}\{\boldsymbol{{\cal S}}_{\cal R}\}    ,
\end{array}
    \label{eq:FieldEqCoarseGrained}
\end{equation}
and where any LHS variable is implicitly understood as long wavelength, i.e.~we've suppressed the the notation ${}^>$.

This whole procedure can now be performed to compute the new right-hand side (RHS) for any dynamical quantity in the ADM formulation and in any of the gauges discussed. Note that the computations are cumbersome and have been checked with \textit{Mathematica} \cite{Mathematica} which was also used to confirm that our previous ${\cal R}$-decompositions were perfectly satisfying the original first-order equation eqs. (\ref{eq:Kcpt_eq}), (\ref{eq:fieldCPT}) and (\ref{eq:ConstraintsCPT}) before applying the window function $W_k$.

Applying the procedure outlined above to all the rest of the ADM equations, we find that they get augmented by Brownian terms as shown below
\begin{equation}
\left \{
\begin{aligned}
 \dot{K}+\beta^i K_{, i}+\alpha^{\mid i}{ }_{\mid i}-\alpha\left({ }^{3} R+K^2\right)&\\
 - M_{Pl}^{-2} \alpha\left(\frac{1}{2} S-\frac{3}{2} \rho\right) =  {\cal F}^{-1}\{\boldsymbol{{\cal S}}_{K}\}, &\\
   \dot{\tilde{K}}_{ij}+2\alpha\tilde{K}_{il}{\tilde{K}}^l{ }_j+\beta^k \tilde{K}_{ij \mid k}-2\beta_i{ }^{\mid k} \tilde{K}_{jk}& \\+ \alpha_{\mid i\mid j}
  -\frac{1}{3} \alpha^{\mid k}{ }_{\mid k}\delta_{ij} 
-\alpha\left({ }^3\tilde{R}_{ij}+\frac{1}{3}K \tilde{K}_{ij}\right) & \\+ M_{Pl}^{-2} \alpha \tilde{S}_{ij}={\cal F}^{-1}\{\boldsymbol{{\cal S}}_{\tilde{K}_{ij}}\},&\\
 \frac{1}{\alpha}\left(\dot{\Pi}+\beta^i \Pi_{\mid i}\right)-K \Pi-\frac{\alpha^{\mid i}}{\alpha} \phi_{\mid i}&\\
 -\phi_{\mid i}^{\mid i}+\frac{d V}{d \phi}={\cal F}^{-1}\{\boldsymbol{{\cal S}}_{\Pi}\},&\\
 { }^{3} R+\frac{2}{3} K^2-\tilde{K}_{i j} \tilde{K}^{i j}-2 M_{Pl}^{-2} \rho = {\cal F}^{-1}\{\boldsymbol{{\cal S}}_{{\cal H}}\}, &\\
 \tilde{K}^j{}_{i \mid j}-\frac{2}{3} K_{\mid i}- M_{Pl}^{-2} {\cal { J }}_i = {\cal F}^{-1}\{\boldsymbol{{\cal S}}_{{\cal M}_j} \}
\end{aligned}\right .
    \label{eq:CoarseGrainedADM}
\end{equation}
where the RHS ${}^>$ superscripts are implicit, the source terms given by
\begin{equation}
\left \{
\begin{aligned}
&  \boldsymbol{{\cal S}}_{K} =  -\varepsilon_1\boldsymbol{{\cal S}}_{{\cal R}} +c.c.,\\ 
& \boldsymbol{{\cal S}}_{\tilde{K}_{ij}} = a^2\varepsilon_1(\frac{1}{3}\delta_{ij}-k^{-2}k_ik_j ) \boldsymbol{{\cal S}}_{{\cal R}}+c.c.,\\
& \boldsymbol{{\cal S}}_{\Pi} = \sqrt{2\varepsilon_1} M_{Pl} \boldsymbol{{\cal S}}_{{\cal R}}+c.c. .
\end{aligned}\right .
    \label{eq:RHSADM}
\end{equation}
and with $\boldsymbol{{\cal S}}_{{\cal R}}$ from eqn.~(\ref{eq:SR}). Importantly, any terms that might contribute to the source terms on the RHS of the constraints \emph{cancel completely} and hence  
\begin{equation}
   \left\{ \begin{aligned}
     & \boldsymbol{{\cal S}}_{{\cal H}}  = 0, \\
    & \boldsymbol{{\cal S}}_{{\cal M}_j}  = 0.
    \end{aligned} \right .
\end{equation}
This a physically appealing result and a consequence of our coarse-graining philosophy: the stochastic noise terms can be thought of as a continuous readjustment of the ``initial value data" at each timestep of the dynamical evolution. When setting up initial conditions in the ADM formalism, all relevant fields (determined by the choice of gauge) must be specified such that the constraints are satisfied (here up to $\mathcal{O}(\boldsymbol{{\cal S}}_{{\cal R}}^2)$). The above equations therefore ensure that this also remains true for each such stochastic readjustment per time step. 

We have performed the computation in all families of gauges previously defined, always finding the same result: the Fourier transforms of the RHS \emph{coincide for all gauges}. We therefore postulate that the above equations must hold for any arbitrary gauge choice even beyond those discussed here. Finally, following the above approach we have also coarse-grained both the more common formulation of Einstein's equations as well as their BSSN incarnation \cite{baumgarte_shapiro_2010}, see Appendix \ref{app:appC}. The results are consistent with eqn.~(\ref{eq:CoarseGrainedADM}). Writing down the BSSN equations is a necessary step to study well-posed GR in numerical relativity.

\subsubsection{Discussion \label{sec:secIVB3}}

Let's first recall that those equations' RHS are only valid for perturbations around a homogeneous background. In particular, without a separate universe approach, we don't provide here any heuristic to account for stochastic backreaction. 

The most striking observation is probably the simplicity and similarity of the RHS terms, which contrasts with the NL LHS. This is completely due to the linear framework in the UV and the fact that all dynamics are encoded in one variable. This is also supported by the numerous null RHS in equations that encode either field definitions or constraints. In particular, the perfect satisfaction of the constraints after coarse-graining is a good sign that our spacetime is physical, i.e. here the horizon crossing is done coherently on the whole time hypersurface but also that we have addressed the insertion of the window. When setting $\Psi_B^> = \Phi_B^>$ one gets a violation of the constraints and a strong gauge-dependence of the RHS, which supports the choice of eqn. (\ref{eq:GaugeInvExplicit>}) for an appropriate coarse-graining.

Another interesting term is that of the anisotropic evolution equation: first-order scalar perturbations source tensorial quantities at higher orders. This is in agreement with previous work \cite{Tomita67,Matarrese98} and is discussed further in section \ref{sec:secIVF}.

Furthermore, the previous equations are valid for four major families of small gauges and any background time slicing (the latter being only a matter of variable change for straight time derivatives to get to $\alpha_b'(t') \neq 1$). This suggests that those equations could be true for any small gauge. At a linear level first, taking, for instance, eqn.(\ref{eq:LinearFieldCoarse}): a 1st order gauge transformation would leave the LHS unchanged and similarly for the RHS as it is written in terms of a gauge-invariant quantity and background time derivatives. This works even if the transformation is a function of the window or ${\cal R}$.
The gauge-invariance of the fully non-linear equations is less obvious. In particular, a gauge transformation will leave any non-linear LHS unchanged but the associated RHS will be consistent only if both the initial and the final gauges are close enough to the homogeneous background one so that we stay within Perturbation Theory when linearising the RHS. This is why there is evidence of small-gauge invariance of our equations at first order.

One needs to emphasize that the long-wavelength approximation has been removed from the equations. This is what allowed us to formulate equations in any gauge by bypassing the gauge-mapping issue in this regime \cite{artigas_hamiltonian_2022}. Note that the long wavelength approximation could still be applied to the choice of the window function if one wants to ensure the complete classicalisation of the crossing modes \cite{burgess_minimal_2022}. However, it seems plausible that the window function can now be turned on much closer to the Hubble radius than in the \textit{$(k=0)$-SUA}, the latter being restricted by the quasi-isotropy assumption \cite{artigas_hamiltonian_2022}. In particular, one can now study safely regime transitions in SI where gradients are critical, as opposed to the main approach \cite{jackson2023}. Switching on the window closer to Hubble crossing also gives less interaction time (and so less higher-order effects) to the UV modes and so strengthens the linear and Gaussian source approximation. Of course, more study is needed before verifying this assertion, something we leave for future work. 

Related to this matter, we finally want to discourage any attempt to go too far away from a dS spacetime. Although it is true that our derivations do not make any slow-roll assumption, using Gaussian sources, linear CPT, and no stochastic backreaction (i.e. full accountancy of UV-IR interactions) might not encode all necessary contributions from UV modes and thus questions the whole story of SI. To go in other regimes, one could study the UV within QFTCS and show the presence of a hierarchy in the correlation functions, order by order.

\subsection{Usual Stochastic Inflation limit  \label{sec:secIVE}}
In this section, we compare our results with eqn.~(\ref{eq:SIrewrite}) of SI \cite{pattison_stochastic_2019}. To achieve this, we need to change the LHS of the $\Pi$ equation to its SUA limit eqn.~(\ref{eq:longGR}) and specify our gauge. 

By fixing $\partial_t{\cal N}=1$, i.e. choosing $t$ as the background efolding ${\cal N}_b$, and $\beta^i=0$, we can write $H^> \equiv -\frac{1}{3}K^> = \frac{1}{\alpha^>}$, and hence $\alpha^>$ is given by the Hamiltonian constraint in the long wavelength limit. Substituting this in the $\Pi$ equation's LHS yields
\begin{equation}
\left \{
\begin{aligned}
 \partial_t {\Pi}^>  & = -3{\Pi}^>-\frac{1}{H^>}\frac{dV}{d\phi}(\phi^>)+ \frac{1}{H}{\cal F}^{-1}\{\boldsymbol{{\cal S}}_{\Pi}\},\\
 {\Pi}^> &=  H^> \partial_t \phi^>, \\
 {H^>}^2 & = \frac{V(\phi^>)}{1-\frac{1}{6M_{Pl}^2}\left(\partial_t \phi^>\right)^2},
\end{aligned}\right .
    \label{eq:SImine}
\end{equation}
which is equivalent to the usual eqn.~(\ref{eq:SIrewrite}) but with our own noise term ${\cal F}^{-1}\{\boldsymbol{{\cal S}}_{\Pi}\}$ at first order. Note that since the ${\cal S}_{\cal R}$ has no apparent Laplacian, we chose to leave it unchanged under the long-wavelength approximation.

Our formalism is now comparable to the literature's SI by looking at $\boldsymbol{\Sigma}$ and $\boldsymbol{{\cal S}_{\cal R}}$ only. On the one hand, ${{\cal S}}_{{\cal R}}$ can be re-written in terms of background efolds slicing ($\alpha_b = H^{-1}$ and $H\partial t = \partial {\cal N}_b$) as
\begin{equation}
    {{\cal S}}_{{\cal R}} = H^2\left ( {\cal R}_k \frac{\partial^2 W_k}{\partial{\cal N}_b^2}+\left[2 \frac{\partial{\cal R}_k}{\partial{\cal N}_b} + (3-\varepsilon_1-\varepsilon_2) {\cal R}_k\right ]\frac{\partial W_k}{\partial{\cal N}_b} \right ).
    \label{eq:MyRHSinEfoldings}
\end{equation}
On the other hand, we can make $\boldsymbol{\Sigma}$ explicit by calculating $\delta \phi$ and $\partial_{{\cal N}_b}\delta \phi$ from the long wavelength limit of our calculus in sec. \ref{sec:secIIIB4} where $\delta{\cal N}^* = 0$. By using the long-wavelength limit\footnote{Note that in \cite{pattison_stochastic_2019}, neglecting the $k^2E$ terms is justified only in SR, USR, and Starobinsky models (and it requires fixing the initial value of $E$, i.e. the remaining d.o.f. of the gauge).}
\begin{equation}
\left \{
\begin{array}{cl}
    \delta\phi   & \simeq \sqrt{2\varepsilon_1}M_{Pl}{\cal R},\\
    \delta\pi &  \simeq \sqrt{2\varepsilon_1}M_{Pl}\left(-\frac{1}{2}\varepsilon_2{\cal R}+\displaystyle\frac{\partial{\cal R}}{\partial{\cal N}_b} \right),
\end{array}\right .
    \label{eq:NoiseUniNlong}
\end{equation}
which confirms to be the same as the spatially flat gauge decomposition. Finally, by substituting this in $\boldsymbol{\Sigma_{\phi}}$ and $\boldsymbol{\Sigma_{\pi}}$ of eqn.~(\ref{eq:SInoise}), one exactly gets $\boldsymbol{\Sigma} =  {\cal F}^{-1}\{\sqrt{2\varepsilon_1}M_{Pl}\boldsymbol{{\cal S}}_{\cal R}\} = {\cal F}^{-1}\{\boldsymbol{{\cal S}}_{\Pi}\}$. 

We just confirmed that our equations give the eqn. (\ref{eq:SIrewrite}) when using the same assumptions.  

\subsection{Stochastic gravitons \label{sec:secIVF}}
As already pointed out, the scalar perturbations have an influence on the non-scalar degrees of freedom through the $\tilde{K}_{ij}$ evolution equation of eqn. (\ref{eq:CoarseGrainedADM}). At second order and later non-perturbative orders, one thus expects scalar-induced and scalar-coupled gravitational waves \cite{Tomita67,Matarrese98}. In that sense, eqn. (\ref{eq:CoarseGrainedADM}) is the first of its kind to provide a stochastic framework including both scalar and tensorial evolutions. This is not surprising as previous studies worked in the long-wavelength limit to drop tensorial dynamics.

In previous stochastic inflation work \cite{Tasinato22}, it is suggested that we should also include the decohered first order gravitons in the stochastic sources. This can be simply added if we stay at linear order for the sources. Indeed, we only need to do the same work as for the scalar but for gravitonic perturbations. This can be done independently and just added to the final scalar result thanks to the scalar-decoupled limit. In practice, this appears to be straightforward as we can build a linear gauge-invariant quantity $h$ by writing the linear tensorial part of the metric in cosmic time as
\begin{equation}
  ds^2 = -dt^2 + a(t)^2(\delta_{ij}+h_{ij})dx^idx^j,  
\end{equation}
with $h^i{}_i=\partial_ih_{ij} = 0$ (traceless and transverse).
At linear order, $K$, $\tilde{K}_{ij}$, ${^3}R$ and ${^3}\tilde{R}_{ij}$ are the only quantities in the EOM inheriting contributions from $h$. This leads to the only equation in (\ref{eq:CoarseGrainedADM}) having $h$ terms at linear order, the linearised $\tilde{K}_{ij}$ equation
\begin{equation}
    -\frac{1}{2}a^2\left[ \partial_t^2 h_{ij}+3H\partial_t h_{ij}-\frac{\nabla^2}{a^2}h_{ij}\right] = 0,
    \label{eq:eqMShreal}
\end{equation}
which in Fourier space appear to be the known Mukhanov-Sasaki equations for the two linear spin-2 components of the gravitons
\begin{equation}
    \left \{
    \begin{aligned}
       & h_{ij}(\vec{k},t)= \epsilon^+_{ij}(\vec{k})h_k^+(t) + \epsilon^\times_{ij}(\vec{k}) h_k^\times(t)\\
       & \ddot{h_k^s}+3H\dot{h_k^s}+ k^2h_k^s = 0, \forall s = +,\times, \\
    \end{aligned}\right .
    \label{eq:eqMSh}
\end{equation}
which can also be initialised by a Bunch-Davies vacuum. 

We now have everything we need to coarse-grain at linear order with the previous window method using
$$h_{ij}^>(\vec{k},t) = W^h_k(t)h_{ij}(\vec{k},t),$$
assuming we want the same window for both polarisations. By coarse-graining eqn. (\ref{eq:eqMSh}) to get the spectrum of sources for eqn. (\ref{eq:eqMShreal}), and promoting the sources to those of the $\tilde{K}_{ij}$ equation, we get to update $\boldsymbol{{\cal S}}_{\tilde{K}_{ij}}$ as
\begin{equation}
\left\{
\begin{aligned}
     \boldsymbol{{\cal S}}_{\tilde{K}_{ij}} & = \boldsymbol{{\cal S}}_{\tilde{K}_{ij}}^{(\phi)}+\boldsymbol{{\cal S}}_{\tilde{K}_{ij}}^{(h)}, \\
 \boldsymbol{{\cal S}}_{\tilde{K}_{ij}}^{(\phi)}  & = a^2\varepsilon_1(\frac{1}{3}\delta_{ij}-k^{-2}k_ik_j ) \boldsymbol{{\cal S}}_{{\cal R}}+c.c.,\\
  \boldsymbol{{\cal S}}_{\tilde{K}_{ij}}^{(h)}  & = -\frac{1}{2}a^2\left(\displaystyle\sum_{s=+,\times}\epsilon^s_{ij}(\vec{k}) \boldsymbol{{\cal S}}_{h}^s\right) +c.c.,\\
  \end{aligned}\right .
\label{eq:eqNewSource}
\end{equation}
  where
  \begin{equation}
\left\{
\begin{aligned}
  \boldsymbol{{\cal S}}_{h,\vec{k}}^s & = \frac{1}{\sqrt{2}}{\cal S}_h(k)\boldsymbol{\alpha}^s_{\vec{k}}+ c.c, \\
  {\cal S}_h(k) & = h_k^s \ddot{W}_k^h + \left[2 \dot{h}_k^s+3Hh_k^s\right] \dot{W}_k^h,\\
  \langle \boldsymbol{\alpha_{\vec{k}_1}}^{s_1}\boldsymbol{\alpha_{\vec{k}_2}}^{s_2*} \rangle_{\mathbb{P}}  & =  \delta^{(3)}(\vec{k}_1-\vec{k}_2)\delta^{s_1s_2}.
\end{aligned}\right .
\label{eq:eqNewSourceWhere}
\end{equation}
The coarse-graining is now complete and appeared to be much easier because of the gauge invariance of the perturbation and the trivial satisfaction of the constraints at linear order. This extension is not without utility because eqn. (\ref{eq:eqNewSource}) shows a competition between scalar and tensor sources, which seems to be in favor of the latter in slow-roll regimes. Tensorial degrees of freedom can now be studied in a non-perturbative framework, most likely numerically in the future. 

\section{Langevin equations in the uniform field gauge \label{sec:secV}}

The stochastic equations of the previous section can be applied in a variety of gauges.  It is now time to extract useful (gauge invariant) quantities such as curvature scalars on certain matter hypersurfaces. In the linear theory, it refers to $\zeta_{gi}$ and ${\cal R}$ defined earlier, curvature perturbations on spatial hypersurfaces of uniform-density and uniform-field respectively. Beyond the first order in CPT, it is possible to construct such non-linear gauge invariant quantities although the literature provides different levels of assumptions \cite{wands_new_2000,rigopoulos_separate_2003,langlois_evolution_2005}. One can in particular define the following ones \cite{rigopoulos_separate_2003,langlois_evolution_2005}
\begin{equation}
\left \{
\begin{aligned}
\frac{1}{6}{\cal \zeta}^{NL}_i & = \partial_i {\cal N}-  \frac{\partial_0{\cal N}}{\partial_0 \rho} \partial_i \rho, \\
\frac{1}{6}{\cal R}^{NL}_i & = \partial_i {\cal N}- \frac{\partial_0{\cal N}}{\partial_0 \phi} \partial_i \phi,
\end{aligned}\right .
    \label{eq:NLgaugeinv}
\end{equation}
where ${\cal N}$ was defined above in sec. \ref{sec:secIIIB4}. It is a gauge-dependent quantity as the determinant is a density-2 tensor. According to \cite{langlois_evolution_2005}, the usefulness of (\ref{eq:NLgaugeinv}) holds even beyond the long wavelength approximation. 

From these variables, it is clear that knowing ${\cal N}$ in uniform-density or uniform-field gauges provides a direct gauge-invariant extraction and this is why the FPTA is required in usual studies where the efolds are not stochastic \cite{vennin_correlation_2015}. In this context and since a gauge-invariant formulation is being proposed in this paper, we propose to apply the uniform field gauge directly to our equations and avoid the FPTA. Note that this has been attempted recently but starting from the literature's usual equations and assumptions \cite{tomberg_numerical_2023}.

We start from our equations (\ref{eq:CoarseGrainedADM}) and set the coordinates such that the field evolves uniformly according to the background dynamics, $\phi(t,\vec{x})=\phi_b(t)$, where $\phi_b(t)$ follows eqn.~(\ref{eq:backgroundEq}) with $\alpha_b = 1$. Hence $\phi$ can act as a clock labeling the 3D spatial hypersurfaces. The spatial coordinates on these spatial slices are fixed such that $\beta^i=0$. Obviously, this gauge choice can only be valid if the background field is non-static. In addition, $\phi$ does not now receive stochastic impulses and the stochastic dynamical variables are ${\cal N}$ and $\alpha$ for which Langevin equations can be derived.   

Leaving a more complete numerical study for future work, we make in this section the long-wavelength approximation for the Hamiltonian constraint, re-writing eqn.~(\ref{eq:SUA_GR})
\begin{equation}
    \frac{6}{\alpha^2}(\partial_t {\cal N})^2 = \frac{2}{M_{Pl}^2}\left(\frac{{{\dot{\phi_b}}}^2}{2\alpha^2}+V(\phi_b)\right),
\end{equation}
by inserting the definition of ${\cal N}$. From this we immediately obtain
\begin{equation}
  \frac{ \partial{\cal N}}{\partial{\cal N}_b} = \displaystyle\sqrt{1+(\alpha^2-1)(1-\frac{1}{3}\varepsilon_1)},
  \label{eq:eqUniFieldefolds}
\end{equation}
by making use of the background Friedman equation. No slow-roll approximation has been made so far.
The second equation we need is given by the field equation which becomes in this gauge an equation for the lapse 
\begin{equation}
  \partial_t \left( \frac{\dot{\phi_b}}{\alpha} \right)= -3\frac{\dot{\phi_b}}{\alpha} \partial_t {\cal N} -\alpha V_{,\phi}(\phi_b) +\alpha\boldsymbol{S}_{\Pi},
\end{equation}
where $\boldsymbol{S}_{\Pi}$ is the Langevin noise found previously in sec. \ref{sec:secIVB}. Note again that any multiplier of $\boldsymbol{S}_{\Pi}$ is left undecided concerning the stochastic backreaction. To get to $\partial_t \alpha = \partial_t f$ where $f(Z_t,t) = \frac{\dot{\phi}_b}{Z_t}$ for  $Z_t = \frac{\dot{\phi}_b}{\alpha}$, the Ito lemma \cite{VanKampen19992, 1994hsmp.book.....G, 2007PhRvE..76a1123L} is used to get 
\begin{equation}
\begin{aligned}
    \partial_t \alpha = &   \frac{\ddot{\phi_b}}{\dot{\phi_b}}\alpha +\frac{\alpha^3}{\dot{\phi_b}^2}\langle\boldsymbol{S}_{\Pi}({\cal N}_b)^2\rangle\\
    & -\frac{\alpha^2}{\dot{\phi_b}} \left (-3\frac{\dot{\phi_b}}{\alpha} \partial_t {\cal N} -\alpha \frac{dV}{d\phi}(\phi_b) +\alpha\boldsymbol{S}_{\Pi}\right ).
    \end{aligned}
    \label{eq:ito}
\end{equation}
We have included the Ito correction to the derivative of a function of a random variable as the second term in the above equation. Note however that the effect of this term is minimal (higher order) and the results of the simulations we describe below are practically unaffected by it. Substituting the field background equation in and the efolds evolution from eqn.~(\ref{eq:eqUniFieldefolds}), changing the time variable by dividing by $H$  and consequently updating the variance of the Wiener process yields
\begin{equation}
\begin{aligned}
    \frac{\partial \alpha}{\partial {\cal N}_b} & = 3\alpha\left(\displaystyle\sqrt{1+(\alpha^2-1)(1-\frac{1}{3}\varepsilon_1)}-1\right) +\alpha^3\langle\boldsymbol{S}({\cal N}_b)^2\rangle\\
    & +(-3+\varepsilon_1+\frac{1}{2}\varepsilon_2)\alpha(\alpha^2-1)+\alpha^3\boldsymbol{S}({\cal N}_b) ,
    \end{aligned}
    \label{eq:eqUniFieldlapse}
\end{equation}
where $\boldsymbol{S} = \boldsymbol{S}_{\Pi}/H\dot{\phi_b}$ is the final stochastic source. In the following, we will keep the coefficient $\alpha^3$ multiplying $\boldsymbol{S}$ as a non-background, stochastic quantity, and won't set it to $1$. Note that when linearized, these equations still match CPT at first order. Equations (\ref{eq:eqUniFieldefolds}) and (\ref{eq:eqUniFieldlapse}) together form a coupled system of stochastic PDEs.  

In the SUA philosophy, it is common to restrain our case to one patch of the universe for the treatment of the noise, i.e. using the previous equation at one given point $\vec{x}_0$ and using only $\boldsymbol{S}({\cal N}_b,\vec{x}_0)$. This is completely justified because the window function derivatives yield suppressed correlations beyond the Hubble scale (e.g. a Heaviside window in Fourier space would lead to a cardinal sinus in real space) \cite{Starobinsky_stochastic_1988}. In this framework, it is easier to solve this system of two coupled Langevin equations or stochastic ODEs.

An analytical solution would require specifying the background dynamics and writing the Fokker-Planck equation to get the PDF. To our knowledge, there is no known way to do that analytically without further approximations such as overdamping, model-dependent simplifications, or higher-order correlations neglection. As a proof of concept, we decided to provide numerical results instead. 

The simulations were realised with the Stratonovich evolver of \textit{ Mathematica} \cite{Mathematica}. The following amplitude for $\boldsymbol{S}$, computed in Appendix \ref{app:appD}, is valid for both slow-roll and ultra-slow regimes (USR)
\begin{equation}
\sqrt{\langle\boldsymbol{S}({\cal N}_b)^2\rangle}\simeq \frac{1}{\sqrt{2\varepsilon_1({\cal N}_b)}}\frac{3}{2\pi}\frac{H({\cal N}_b)}{M_{Pl}}.
\end{equation}
The main difference with usual amplitudes is the $\sqrt{\varepsilon_1}$ in the denominator. This comes from the fact that the lapse equation depends on the amplitude of ${\cal R}$ and not the usual Mukhanov-Sasaki variable ($a\sqrt{2\varepsilon_1}M_{Pl}\times{\cal R}$) to which field equations are sensitive to. Of course, the $\varepsilon_1\rightarrow 0$ limit is problematic but this is purely a gauge artifact as already noted earlier when defining our gauge and as explained in \cite{Kinney05}.  For the simulation, we choose to evaluate this quantity with the background evolution and so to neglect the stochastic backreaction for simplicity, as opposed to the $\alpha^3$ factor acting on $\boldsymbol{S}$.

\begin{figure*}
    \centering
\includegraphics[width=0.9\textwidth]{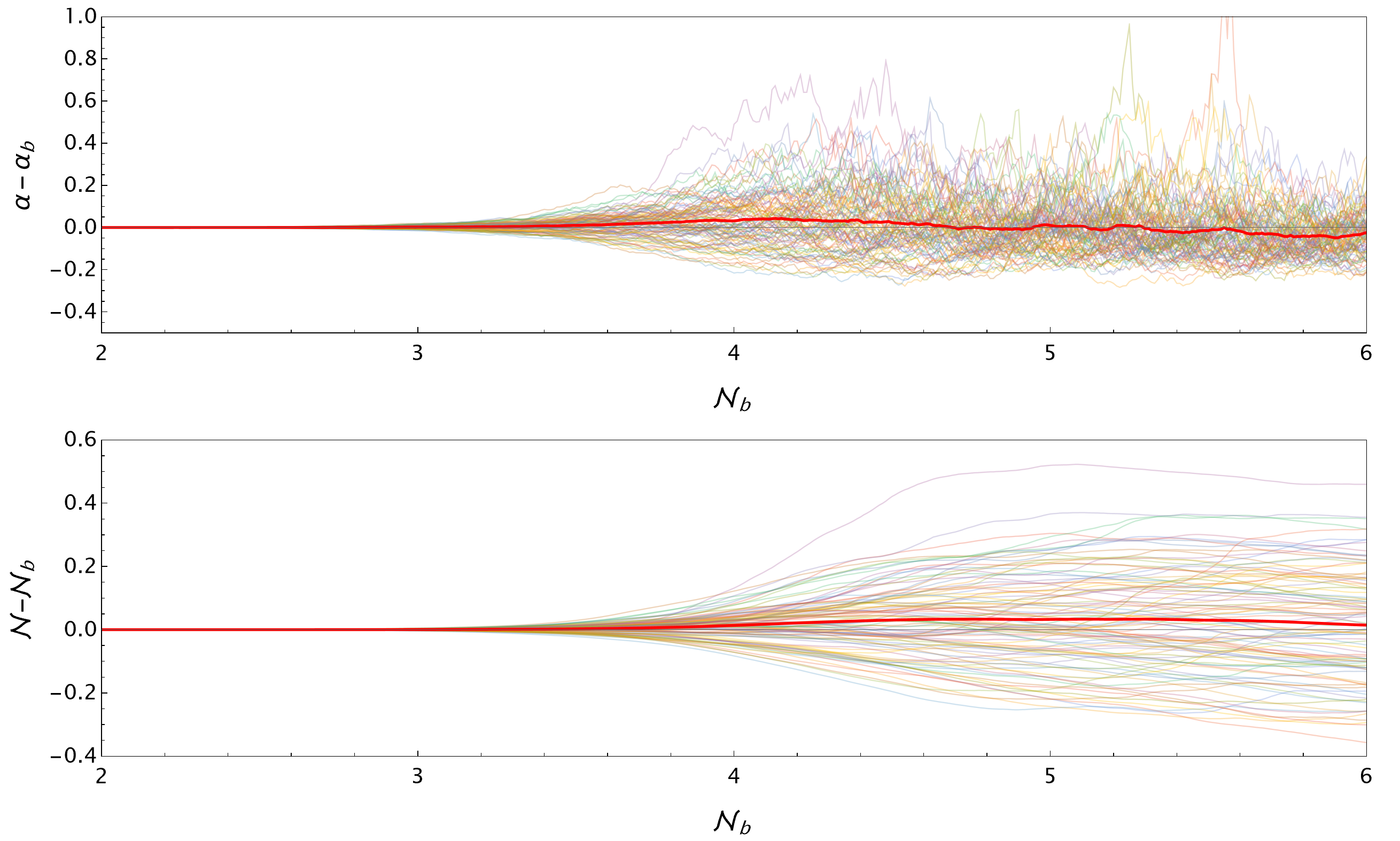}
    \caption{A scalar field undergoing ultra slow-roll dynamics in uniform field gauge, see eqn.(\ref{eq:eqUniFieldlapse}), (\ref{eq:eqUniFieldefolds}). The noise is active between $2$ and $6$ efolds. 100 of 50000 stochastic paths are shown for the lapse (top) and efold (bottom) differences to background. Red curves provide the mean difference over the 100 paths. The field has been initialised at $\phi_b^* = 10$ with an efolds velocity $\pi_b^*=d\phi_b/d{\cal N}_b^* = -2.0001$ in a potential such that $V(\phi)=V_0$ until $\phi_c = 9.056$, where after that $V(\phi)=V_0 [3 (\phi/\phi_c)^2-2(\phi/\phi_c)^3]$, which stops the exponential slowing of the field.  Note that $V_0$ is set to satisfy the Friedman constraint equation assuming $H^* = 10^{-5}$ initially. Quantities are expressed in $M_{Pl}= c = 1$ units.}
    \label{fig:fig1}
\end{figure*}

\begin{figure*}
    \centering
\includegraphics[width=0.65\textwidth]{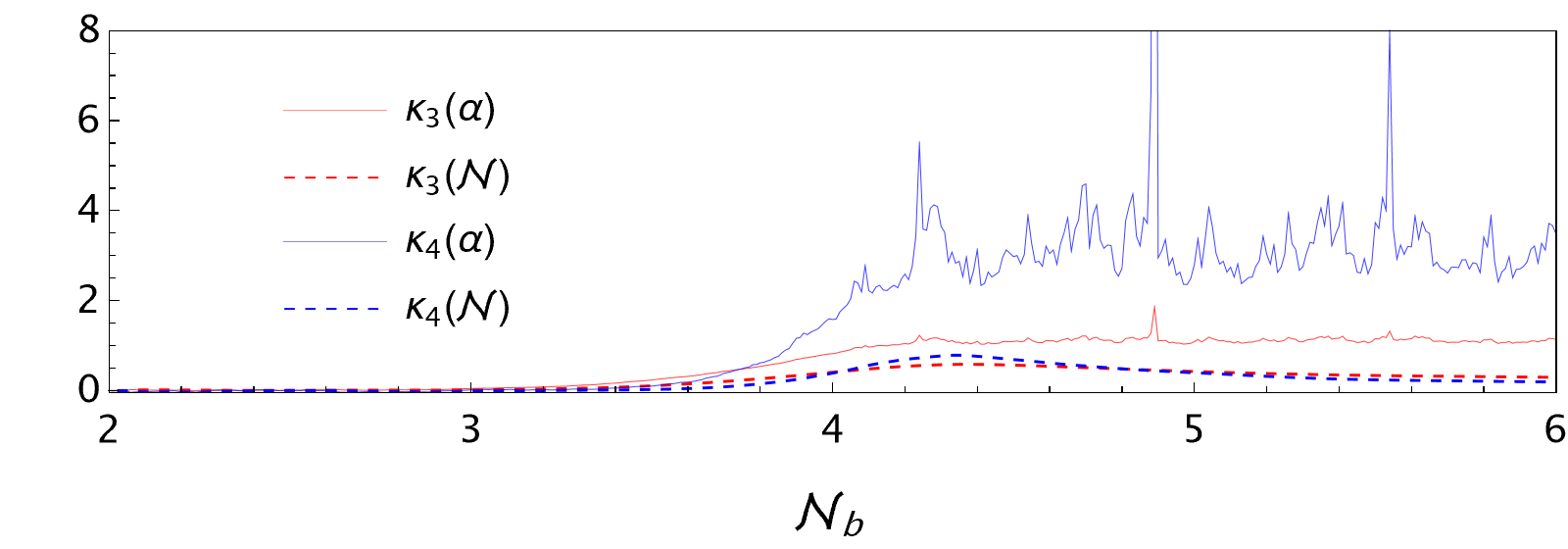}
    \caption{Skewness and kurtosis excess in time for 50000 stochastic paths of the lapse and efold differences to background respectively. See Figure \ref{fig:fig1} for the parameters of the simulation.}
    \label{fig:fig2}
\end{figure*}

Figures \ref{fig:fig1} and \ref{fig:fig2} show 50.000 paths starting from uniform lapse and efolds. $0.01$-efolds steps were used in the numerical scheme. Without worrying about the realism of our model of Inflation concerning observational constraints, we considered a plateau Inflation followed continuously by a 3rd order polynomial slope of the form $V(\phi)\propto[3 (\phi/\phi_c)^2-2(\phi/\phi_c)^3]$. Initial position and momentum together with the start of the slope are fine-tuned\footnote{Note that this precise tuning depends on our time resolution ($0.01$ efolds)} for two reasons: we want a decent amount of dispersion and non-Gaussianity by having the field almost stopped on the plateau, but not too much as the gauge definition makes the simulation crash if the velocity is too low. Constraints also need to be satisfied. It appears that USR is reached from 2 efolds onwards ($\varepsilon_1 \ll1$, $\varepsilon_2 \simeq 6$) which is why the noise is only turned on at this time. When the slope is reached at $\simeq 3.04$ efolds, the velocity starts to increase extremely slowly, still within the USR regime. 

\begin{figure*}
\includegraphics[width=0.78\textwidth]{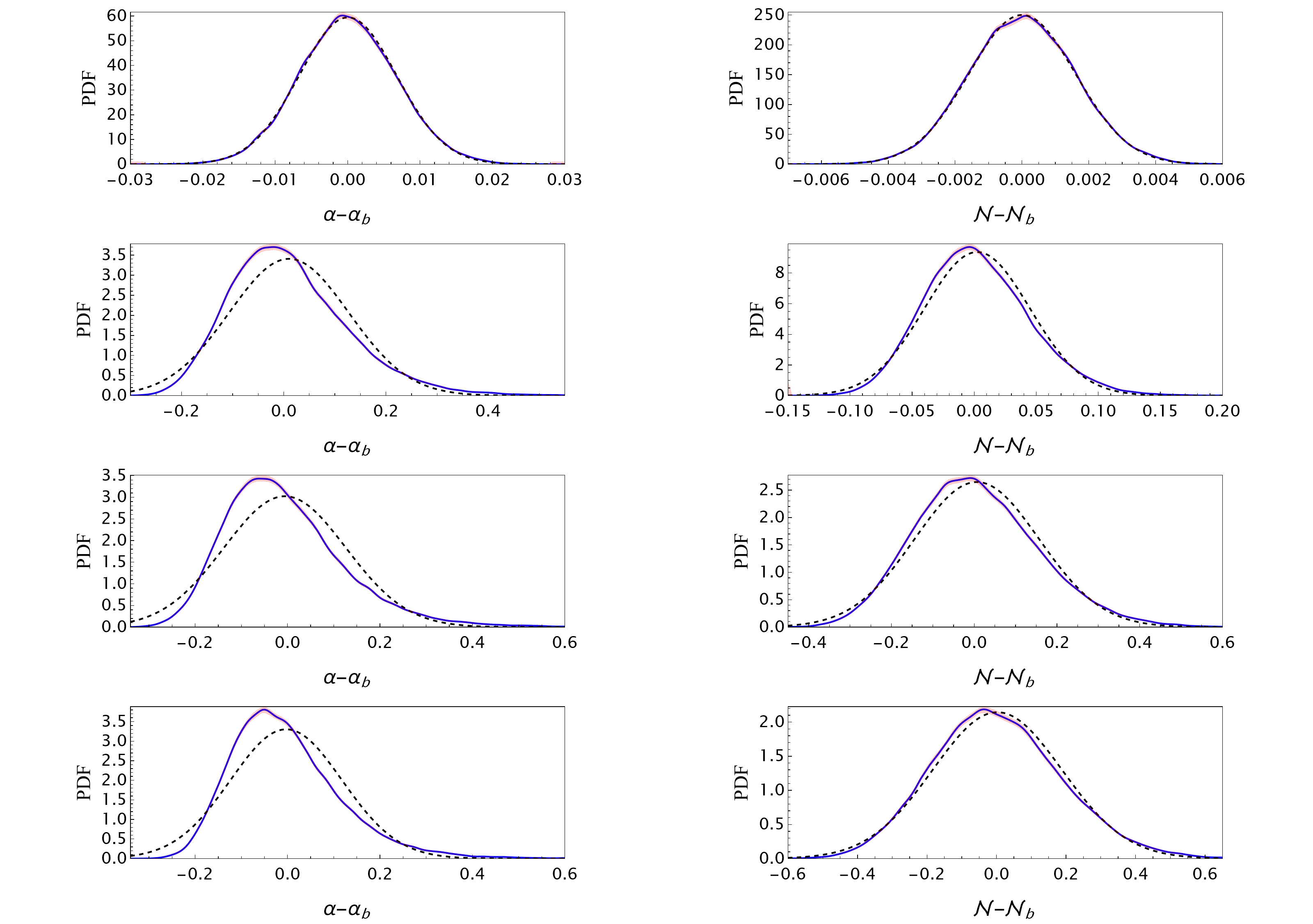}
    \caption{PDFs (blue) of a scalar field undergoing USR dynamics in uniform field gauge, see Figure \ref{fig:fig1} for parameters of the simulation. 50 000 paths were used for Kerner Density Estimation to probe the PDFs at five background efold times (${\cal N}_b=3,4,5,6$ from top to bottom) for both the lapse (left) and the efold (right) differences to the background, against normal PDFs with the same first two moments (black dashed). $95\%$ confidence intervals (red) of these pdf estimators were calculated using bootstrapping.}
    \label{fig:fig3}
\end{figure*}

\begin{figure*}
\includegraphics[width=0.78\textwidth]{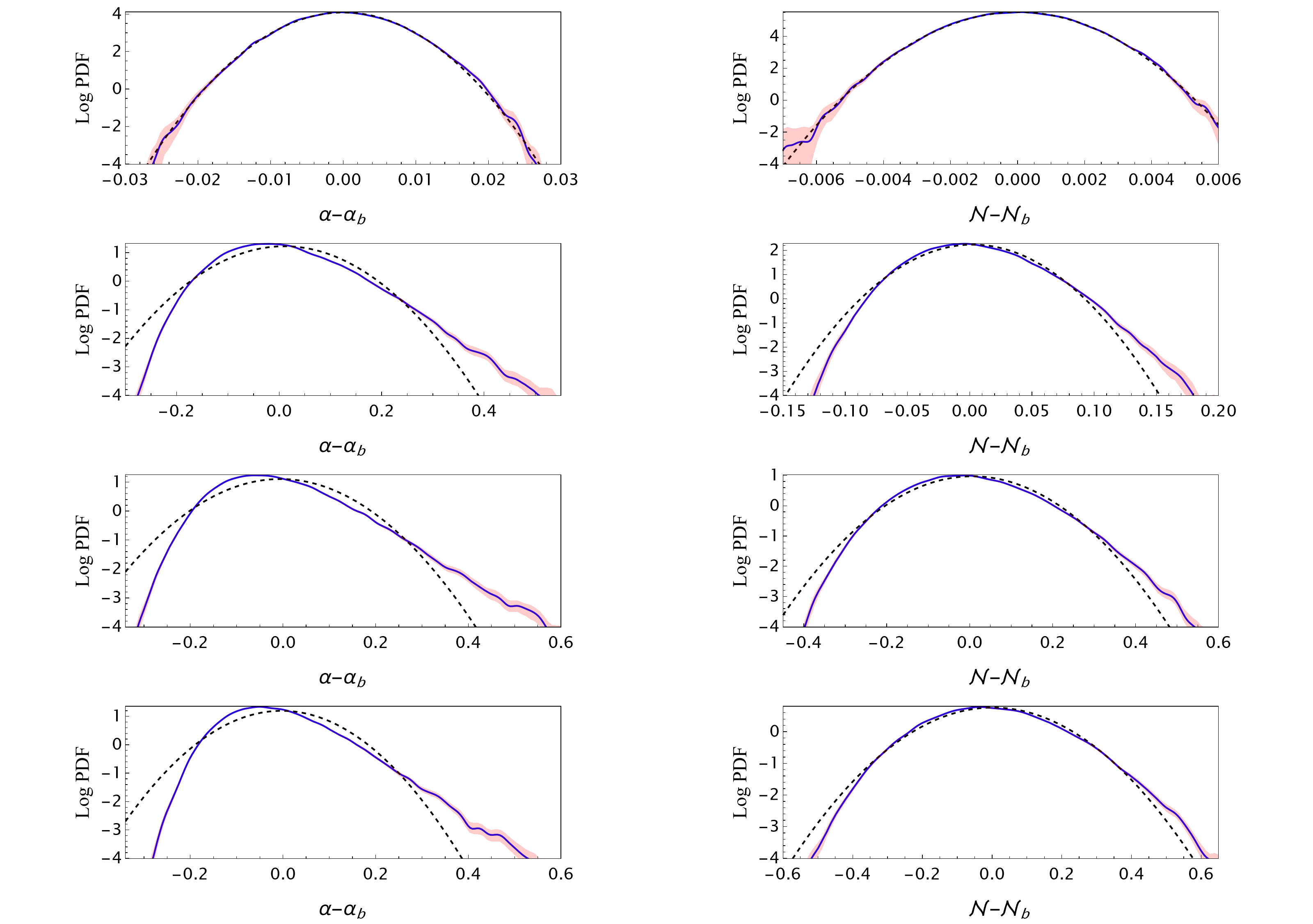}
    \caption{Log PDFs of a scalar field undergoing USR dynamics in uniform field gauge, see Figure \ref{fig:fig3}.}
    \label{fig:fig4}
\end{figure*}

It is well known that a flat potential or a transition can leave strong non-Gaussian imprints on perturbations such as exponential tails \cite{ezquiaga_exponential_2020}. This is actually what we confirm here in Figures \ref{fig:fig3} and \ref{fig:fig4}: both the lapse and the efolds get an exponential tail on the plateau between $3$ (end of plateau) and $4$ efolds, which can be confirmed by diverse fittings. This non-Gaussianity can also be tracked in time by looking at the skewness and the kurtosis in Figure \ref{fig:fig2}.

When it comes to the second phase - and any analytical attempts would probably fail to describe it fully, things eventually stabilise along the slope. From these figures, it becomes clearer that the lapse acts as an extremely non-Gaussian efolds' momentum. This implies that very strong non-Gaussian changes are given to the efolds until stopping and eventually the efolds distribution's non-Gaussianity stabilizes later to a lower remnant level\footnote{A study of the PDF's small tail would be possible using the constrained stochastic formalism \cite{Tokeshi23} or importance sampling \cite{Jackson22}.} when all realisations are in the same regime. We have checked that switching the noise terms off at $6$ efolds makes the lapse come back to its attractor $\alpha_b =1$ and that the efolds, which are meant to describe ${\cal R}^{NL}$, are indeed conserved. If the noise was frozen later on this potential, one would see that leaving the USR phase makes the lapse and the efolds back closer to Gaussianity. This is because the PDF in real space is not a good estimator: adding many Gaussian contributions lowers the relative non-Gaussianity. However the Non-Gaussianity we produced is still imprinted by the end of our simulation, in particular on the statistics of the scales which crossed the Horizon before 6 efolds. This advocates for the necessity to look at quantities such has $n$-spectra or coarse-grained PDFs when looking at data.

It is important to mention that the simulation has also been run by setting $\alpha^3$ to $\alpha_b^3 = 1$. Unusual left-skewed PDFs have been generated and highlight the importance of stochastic backreaction in certain cases. Here it is critical because of the non-perturbative behavior of the lapse and the efolds. In particular, this term is a good barrier to reaching an unphysical $\alpha=0$ with stochastic kicks, without adding a prior when solving the Fokker-Planck equation.

What is to be remembered from these equations and simulation is their potential: having coarse-grained but satisfied constraints and an idea of what the stochastic amplitude is, can help us skip the tedious and approximate stochastic $\Delta {\cal N}$  and FPT formalisms by writing it all directly in the right gauge. 

However, sticking to the literature, the SUA was used and so the validity is far from the crossing scale. This allows us to reduce to a simple GR framework where we don't need a full Numerical Relativity code which probably does not exist in such a gauge.  Furthermore, the validity of the long-wavelength approximation is questionable when perturbations become smaller in magnitude than gradient corrections. This could be the case here for the lapse. For these reasons, traceless modes and other terms should be fully accounted for and could stop those perturbations from vanishing completely. We thus emphasize here that it is necessary to run a full NR code even if it takes to use an NR gauge. In that sense, this section mainly aims at illustrating how this new framework might compete with the FPT formalism using the same assumptions for the evolution.

\twocolumngrid

\section{Conclusions and discussion\label{sec:secVI}}

In this work we have examined the formulation of SI in full General Relativity, dropping most of the approximations made historically such as the long wavelength approximation and the corresponding reduction of gravitational field variables, such as anisotropic degrees of freedom or the momentum constraint, that have been integral to existing versions of the SUA. We have also addressed the issue of time-slicing choice; the choice of lapse and shift in the language of ADM. The only essential approximation kept here is the requirement that CPT is enough to compute the noise source terms, a requirement that could be possibly lifted by retaining higher orders in CPT. Although we have been invariably discussing `stochastic' noise terms, the coarse-graining behind our derivations could have been made at any scale, provided one was content with operator-valued source terms. Of course, the classicalization of cosmological perturbations is highly convenient, and it is this requirement that sets the IR scale to be placed at some slightly above the Hubble radius.        

Returning to the core of 3+1 general relativity, which is essentially the evolution of one temporal 3D-hypersurface to the next, we have proposed a method to coarse-grain the linear theory in a gauge-invariant way using its only degree of freedom, here chosen to be $\mathcal{R}$ the comoving curvature perturbation. We have validated our procedure for several of the most common gauges used in cosmology or numerical relativity, finding that the choice of any of them always results in the same source terms, at least at the linear level. Linking the resulting Langevin terms to their counterpart non-linear equations and adding the treatment of stochastic gravitonic sources, we have provided the first complete set of GR equations for Stochastic Inflation. Looking at their form and exploring alternative gauges has offered strong evidence for our postulate that they are indeed gauge-invariant in that any gauge choice would provide identical results. 

From our equations, we were able to recover the limits where existing SI equations apply. We were also able to go beyond the usual approaches to demonstrate broader applications; our example showed that our formalism could obtain results for the stochastic dynamics of the e-folds $\mathcal{N}$ directly in the uniform field gauge, without shifting 3D-hypersurfaces as required in the stochastic $\Delta {\cal N}$ formalism, for the same SUA assumptions normally made in the existing literature.

We want to highlight the potential of such results. The possible applications are numerous and most notably a key focus of our future work will be numerical. The present article provides all the tools needed for sourcing a full 3+1 numerical relativity code (most of which are running with BSSN equations) with stochastic perturbations. Note that this includes the Langevin terms but also the initial conditions (see sec. \ref{sec:secIIIB4}) which constitute the main challenge of numerical relativity. The present systematic treatment gives the opportunity to quantitatively study the nonlinear evolution of inhomogeneities, taking forward previous work which considered inflationary initial conditions in a variety of contexts \cite{aurrekoetxea_oscillon_2023,clough_scalar_2017, aurrekoetxea_cttk_2022, giblin_jr_preheating_2019,joana_cosmic_2022}.
Full GR simulations of super-Hubble dynamics during inflation should provide insights about the nonlinear generation of higher-order correlators. This is important because currently there is no alternative to QFTCS methods except Stochastic Inflation, which has to date traded greater scope on super-Hubble scales in exchange for numerous other approximations. 

Finally, we note that there remain many further extensions to be considered for Stochastic Inflation. For instance, our methods should be compatible with a greater range of inflationary scenarios such as multiple fields \cite{rigopoulos_non-linear_2005}, accounting for anisotropic sources from these, or modified and higher energy theories of gravity in the context of the EFT of Inflation \cite{Clifford08}. Note that these would require extra-work to find well-posed formulations, enabling numerical solving. It would also be interesting to investigate non-quasi-dS spacetimes, though this would either require rigorous justifications or higher-order perturbations and statistics. With or without quasi-dS scenarios, the incorporation of higher-order effects also lacks a full GR framework. 
As stated in \cite{morikawa_dissipation_1990, moss_effective_2017,Gorbenko19,celoria_2021} in fixed dS spacetimes, we believe that the most rigorous approach would begin from a path integral approach rather than the EOM, however, efforts should be made towards a full GR framework \cite{prokopec_path_2010}. In particular, the coarse-graining approximations would be under control and this could enable the incorporation of quantum loops, clarifying the validity of Starobinsky's approximation with all gravitational degrees of freedom. This is left for future work.

\begin{acknowledgments}
Y.L. thanks all individuals with whom fruitful discussions and debates took place, for instance, E. Florio, C.McCulloch, S.Mishra, E. Pajer, D-G. Wang, T. Colas, B. Sherwin and P. Benincasa. Y.L. is supported by the STFC DiS-CDT scheme and the Kavli Institute for Cosmology, Cambridge.  G.I.R. would like to thank T. Prokopec and I. Moss for many discussions on stochastic inflation and D. Cruces and C. Germani for clarifying their recent work. E.P.S.S. acknowledges funding from STFC Consolidated Grant No. ST/P000673/1.
\end{acknowledgments}

\appendix

\section{Background quantities \label{app:appA}}
Useful background quantities can be expressed as 
\begin{equation}
\left \{
\begin{array}{ll}
 {\dot{\phi_b}}^2 & = 2\varepsilon_1 M_{Pl}^2 H^2,\\
 \dot{H} & = -\varepsilon_1 H^2, \\
 \ddot{H} & = 2\varepsilon_1(1-\varepsilon_2/2)H^3, \\
\ddot{a} & = (1-\varepsilon_1)a H^2, \\
\ddot{\varepsilon}_1 & = \varepsilon_1 \varepsilon_2 (\varepsilon_1+\varepsilon_2+\varepsilon_3) H^2, \\
V_{,\phi}(\phi_b) & = \sqrt{2\varepsilon_1}(-3+\varepsilon_1+\varepsilon_2/2) M_{Pl} H^2, \\
V_{,\phi\phi}(\phi_b) & = -\frac{1}{4}[8\varepsilon_1^2+2\varepsilon_1(-12+5\varepsilon_2)\\
&+\varepsilon_2(-6+\varepsilon_2+2\varepsilon_3)] H^2, \\
\end{array}\right .
\end{equation}
where $ \varepsilon_{i+1} = -H^{-1}d_t \ln \varepsilon_i$
and $\varepsilon_0$ is $H$. The sign of $\dot{\phi}_b$ is implicitly positive by convention.

\section{Degrees of freedom in small generalised synchronous gauges  \label{app:appB}}

To understand this freedom, let us take a look at the effect of the following gauge transformation
\begin{equation}
\left\{
    \begin{aligned}
        \tilde{x}^i & \longrightarrow x^i+\partial^i\lambda,  \\
\tilde{t}& \longrightarrow t + \zeta,
    \end{aligned}\right .
\end{equation}
where $\lambda(t,\vec{x}) = f(\vec{x})-g(\vec{x})(t-t^*)/a(t)^2 $ and $\zeta(\vec{x}) = -g(\vec{x})$.

Under this transformation the lapse and the shift perturbations are unchanged
\begin{equation}
\left\{
    \begin{aligned}
\Psi \longrightarrow &  \Psi-\dot{\zeta} = \Psi, \\
B  \longrightarrow & B + \zeta/a^2-\dot{\lambda} \\
& = B -g/a^2+g/a^2 = B.
    \end{aligned}\right .
\end{equation}
The conclusion is different for $E$ and $\chi$
\begin{equation}
\left\{
    \begin{aligned}
E(t, \vec{x})\longrightarrow &  E(t,\vec{x}) - \lambda(t, \vec{x}) \\
& =E(t,\vec{x})- f(\vec{x})+g(\vec{x})(t-t^*)/a(t)^2,\\
\chi(t, \vec{x}) \longrightarrow &  \chi(t, \vec{x})  -\zeta(t, \vec{x}) = \chi(t, \vec{x})+g( \vec{x}),
    \end{aligned}\right .
\end{equation}
which evaluated at $t^*$ shows that $f(\vec{x})$ and $g(\vec{x})$ can be chosen to set the gauge with any desired value of $\chi(t^*,\vec{x})$ and $E(t^*,\vec{x})$. Thus, a small 3+1 slicing still has some gauge freedom.

\section{Other GR-Langevin equations \label{app:appC}}
\subsection{Einstein-Langevin equations}
Similarly, the coarse-graining of Einstein's equation was also successful. The Einstein-Langevin equation for SI are
\begin{equation}
\left \{
\begin{aligned}
G_{00}-M_{Pl}^{-2}T_{00} =& 0,  \\
G_{0i}-M_{Pl}^{-2}T_{0i} = & 0 ,\\
G_{ij}-M_{Pl}^{-2}T_{ij} = & {\cal F}^{-1}\{\boldsymbol{{\cal S}}_{ij}\}  ,\\
\end{aligned}\right .
    \label{eq:EinsteinCoarseGrained}
\end{equation}
where 
\begin{equation}
\boldsymbol{{\cal S}}_{ij}  = -a^2\varepsilon_1(\delta_{ij}-k^{-2}k_ik_j ) \boldsymbol{{\cal S}}_{{\cal R}}+c.c..
    \label{eq:RHSEinstein}
\end{equation}

\subsection{BSSN-Langevin equations}
When it comes to numerical simulations, it is usually convenient to reformulate any system of PDEs into a well-posed one with first-order equations only. BSSN equations are now the common equations for these purposes \cite{BSSN_SN,BSSN_BS}.

In this formalism, the metric writes
\begin{equation}
    d s^2=-\alpha^2 d t^2+\gamma_{i j}\left(d x^i+\beta^i d t\right)\left(d x^j+\beta^j d t\right),
        \label{eq:BSSNmetric}
\end{equation}
where $\alpha$ and $\beta^i$ are the lapse and shift, gauge parameters, which is the same as our ADM metric but with $\beta^i \longrightarrow-\beta^i$. The induced metric is decomposed thanks to a conformal factor $X$ as
\begin{equation}
\gamma_{i j}=\frac{1}{X} \tilde{\gamma}_{i j}, \operatorname{det} \tilde{\gamma}_{i j}=1, X=\left(\operatorname{det} \gamma_{i j}\right)^{-\frac{1}{3}},
    \label{eq:ConformalFactor}
\end{equation}
The extrinsic curvature is decomposed into its trace $K=$ $\gamma^{i j} K_{i j}$, and its conformally rescaled traceless part $\tilde{\gamma}^{i j} \tilde{A}_{i j}=0$ as
\begin{equation}
K_{i j}=\frac{1}{X}\left(\tilde{A}_{i j}+\frac{1}{3} K \tilde{\gamma}_{i j}\right).    
    \label{eq:ConformalK}
\end{equation}
Finally, an intermediary quantity is defined to break the equations into first-order ones: the conformal connections defined as  $\tilde{\Gamma}^i=\tilde{\gamma}^{j k} \tilde{\Gamma}_{j k}^i$ where $\tilde{\Gamma}_{j k}^i$ are the Christoffel symbols associated with the conformal metric $\tilde{\gamma}_{i j}$. 
As a summary, NR consists of evolving the 7 quantities $X$, $\tilde{\gamma}_{ij}$, $K$, $\tilde{A}_{ij}$, $\tilde{\Gamma}^i$ and $\phi$, $\Pi$.

As it is just a re-writing of ADM, the same reasoning as performed in sec. \ref{sec:secIVB} can be used with \textit{Mathematica} for both the linearization and the coarse-graining. The associated Langevin BSSN equations for SI are
\begin{equation}
\left\{
\begin{aligned}
 \partial_t X -\frac{2}{3} X \alpha K+\frac{2}{3} X \partial_k \beta^k-\beta^k \partial_k X = {\cal F}^{-1}\{\boldsymbol{{\cal S}}_{X}^{BSSN}\},&\\
 \partial_t \tilde{\gamma}_{i j}+2 \alpha \tilde{A}_{i j}-\tilde{\gamma}_{i k} \partial_j \beta^k-\tilde{\gamma}_{j k} \partial_i \beta^k &\\
 \quad+\frac{2}{3} \tilde{\gamma}_{i j} \partial_k \beta^k-\beta^k \partial_k \tilde{\gamma}_{i j} ={\cal F}^{-1}\{\boldsymbol{{\cal S}}_{\tilde{\gamma}_{ij}}^{BSSN}\}, &\\
  \partial_t K+\gamma^{i j} D_i D_j \alpha-\alpha\left(\tilde{A}_{i j} \tilde{A}^{i j}+\frac{1}{3} K^2\right) &\\
-\beta^i \partial_i K-4 \pi \alpha(\rho+S) = {\cal F}^{-1}\{\boldsymbol{{\cal S}}_{K}^{BSSN}\}, &\\
 \partial_t \tilde{A}_{i j}-X\left[-D_i D_j \alpha+\alpha\left(R_{i j}- M_{Pl}^{-2}\alpha S_{i j}\right)\right]^{{TF}} &\\
-\alpha\left(K \tilde{A}_{i j}-2 \tilde{A}_{i l} \tilde{A}_j^l\right)
-\tilde{A}_{i k} \partial_j \beta^k-\tilde{A}_{j k} \partial_i \beta^k &\\
+\frac{2}{3} \tilde{A}_{i j} \partial_k \beta^k-\beta^k \partial_k \tilde{A}_{i j} = {\cal F}^{-1}\{\boldsymbol{{\cal S}}_{\tilde{A}_{ij}}^{BSSN}\},&\\
 \partial_t \tilde{\Gamma}^i-2 \alpha\left(\tilde{\Gamma}_{j k}^i \tilde{A}^{j k}-\frac{2}{3} \tilde{\gamma}^{i j} \partial_j K-\frac{3}{2} \tilde{A}^{i j} \frac{\partial_j X}{X}\right) &\\
+2 \tilde{A}^{i j} \partial_j \alpha-\beta^k \partial_k \tilde{\Gamma}^i 
-\tilde{\gamma}^{j k} \partial_j \partial_k \beta^i-\frac{1}{3} \tilde{\gamma}^{i j} \partial_j \partial_k \beta^k &\\
-\frac{2}{3} \tilde{\Gamma}^i \partial_k \beta^k+\tilde{\Gamma}^k \partial_k \beta^i+ 2 M_{Pl}^{-2}\alpha \tilde{\gamma}^{i j} S_j = {\cal F}^{-1}\{\boldsymbol{{\cal S}}_{\tilde{\Gamma}}^{BSSN}\}, &\\
 \partial_t \phi-\alpha \Pi-\beta^i \partial_i \phi = 0, &\\
 \partial_t \Pi-\beta^i \partial_i \Pi-\alpha \partial_i \partial^i \phi-\partial_i \phi \partial^i \alpha &\\
-\alpha\left(K \Pi-\gamma^{i j} \Gamma_{i j}^k \partial_k \phi-\frac{d V}{d \phi}\right)= {\cal F}^{-1}\{\boldsymbol{{\cal S}}_{\Pi}^{BSSN}\},  &\\
         \mathcal{H}=R+K^2-K_{i j} K^{i j}- 2 M_{Pl}^{-2}\rho = {\cal F}^{-1}\{\boldsymbol{{\cal S}}_{{\cal H}}^{BSSN}\}, & \\
        \mathcal{M}_i=D^j\left(\gamma_{i j} K-K_{i j}\right)- M_{Pl}^{-2}S_i ={\cal F}^{-1}\{ \boldsymbol{{\cal S}}_{{\cal M}}^{BSSN}\}, &
    \end{aligned}\right .
        \label{eq:BSSNCoarseGrained}
\end{equation}
where the RHS Fourier transforms are similar to ADM's after explicit calculus
\begin{equation}
\left \{
\begin{aligned}
& \boldsymbol{{\cal S}}_{X}^{BSSN} = 0,\\
& \boldsymbol{{\cal S}}_{\tilde{\gamma}_{ij}}^{BSSN} = 0, \\
&  \boldsymbol{{\cal S}}_{K}^{BSSN} =\boldsymbol{{\cal S}}_{K}+c.c.,\\
&  \boldsymbol{{\cal S}}_{A_{ij}}^{BSSN}  =a^{-2} \boldsymbol{{\cal S}}_{\tilde{K}_{ij}}+c.c.,\\
& \boldsymbol{{\cal S}}_{\Pi}^{BSSN} = \boldsymbol{{\cal S}}_{\Pi}+c.c.,\\
& \boldsymbol{{\cal S}}_{\tilde{\Gamma}_i}^{BSSN} =0, \\
& \boldsymbol{{\cal S}}_{{\cal H}}^{BSSN}  =  0, \\
& \boldsymbol{{\cal S}}_{{\cal M}_j}^{BSSN}  = 0.
\end{aligned}\right .
    \label{eq:RHSBSSN}
\end{equation}

\section{(U)SR noise amplitude \label{app:appD}}
The Fourier amplitude of $\boldsymbol{S}$ is exactly the same as for ${\cal S}_{\cal R}/H^2$ in efolding time (see eqn. (\ref{eq:MyRHSinEfoldings}))
\begin{equation}
    {{ S}}_k = {\cal R}_k \frac{\partial^2 W_k}{\partial{\cal N}_b^2}+\left[2 \frac{\partial{\cal R}_k}{\partial{\cal N}_b} + (3-\varepsilon_1-\varepsilon_2) {\cal R}_k\right ]\frac{\partial W_k}{\partial{\cal N}_b}. 
    \label{eq:eqSk}
\end{equation}
We choose to work with a Heaviside window
\begin{equation}
    W_k = \Theta(\sigma aH-k),
\end{equation}
far enough from Hubble crossing thanks to $\sigma \ll 1$ and the derivative of which is the Dirac distribution
\begin{equation}
    \partial_{{\cal N}_b}W_k = \partial_{{\cal N}_b} (\sigma aH)\delta(\sigma aH-k).
\end{equation}
The second derivative needs to be treated within distribution theory, which is why we choose to write
\begin{equation}
    {\cal F}^{-1}\{{\cal R}_k \partial^2_{{\cal N}_b} W_k\} =  -{\cal F}^{-1}\{\partial_{{\cal N}_b}{\cal R}_k \partial_{{\cal N}_b} W_k\}.
\end{equation}
It is then allowed to take eqn.(\ref{eq:eqSk}) as
\begin{equation}
    {{ S}}_k = \left[\partial_{{\cal N}_b}{\cal R}_k + (3-\varepsilon_1-\varepsilon_2) {\cal R}_k\right ] \partial_{{\cal N}_b} W_k.
    \label{eq:eqSk2}
\end{equation}
The solution of ${\cal R}_k$ can be found by solving eqn.(\ref{eq:R_evol_eq}) in the case of slow-roll and ultra-slow roll. It turns out that the solution is identical despite different equations \cite{Kinney05} and written in conformal time $\tau$ as
\begin{equation}
    {\cal R}_k(\tau) = \frac{H}{\sqrt{4\varepsilon_1 M_{Pl}^2k^3}}(1+i k \tau)e^{-ik\tau}.
\end{equation}
It is now convenient to verify that
\begin{equation}
    \partial_{{\cal N}_b}{\cal R}_k \delta(\sigma aH-k) = \left(\frac{1}{aH} \partial_{\tau}{\cal R}_k \right )_{-k\tau=\sigma}= -\frac{\sigma^2}{\sqrt{4\varepsilon_1 M_{Pl}^2k^3}}e^{i\sigma},
\end{equation}
which is negligible compared to the ${\cal R}_k$ term in eqn.(\ref{eq:eqSk2}) in the $\sigma\ll1$ limit. 

The amplitude of the Gaussian noise is found by performing the inverse Fourier transform of the spectrum \cite{Starobinsky_stochastic_1988} with spherical invariance

\begin{equation}
\begin{aligned}
    \sqrt{\langle\boldsymbol{S}({\cal N}_b)^2\rangle} & = \left(\frac{|(3-\varepsilon_1-\varepsilon_2) {\cal R}_{\sigma a H}|^2}{6\pi^2}\frac{d}{d{\cal N}_b}(\sigma a H)^3 \right)^{\frac{1}{2}}\\
    & \simeq \frac{1}{\sqrt{2\varepsilon_1}}\frac{3}{2\pi}\frac{H({\cal N}_b)}{M_{Pl}},
\end{aligned}
\end{equation}

using $|{\cal R}_{\sigma aH}| \simeq \frac{H}{\sqrt{4\varepsilon_1 M_{Pl}^2k^3}}$ and in the limit where $\varepsilon_1 \simeq 0 $ and $\varepsilon_2 \simeq 0$ (SR)  or $\varepsilon_2 \simeq 6$ (USR).

\bibliography{Stochastic_Inflation}
\end{document}